\documentclass[preprint,12pt]{elsarticle}
\journal{Annals of Physics}
\RequirePackage{bm,,amsfonts,lineno,hyperref,amsmath,amssymb,microtype,float}
\usepackage{color}

\begin{document}

\begin{frontmatter}

\title{Strange stars in Krori-Barua spacetime under $f(R,T)$ gravity}

\author{Suparna Biswas$^a$, Shounak Ghosh$^a$, Saibal Ray$^{b}\footnote{$^*$Corresponding author.\\
{\it E-mail addresses:} sb.rs2016@physics.iiests.ac.in (SB), shounak.rs2015@physics.iiests.ac.in (SG), saibal@associates.iucaa.in (SR), rahaman@associates.iucaa.in (FR), dean.fa@iiests.ac.in (BKG).}$, Farook Rahaman$^c$, B.K. Guha$^a$}

\address{$^a$Department of Physics, Indian Institute of Engineering Science and Technology, Shibpur, Howrah, West Bengal, 711103, India\\
$^b$Department of Physics, Government College of Engineering and Ceramic Technology, Kolkata 700010, West Bengal, India \& Department of Natural Sciences, Maulana Abul Kalam Azad University of Technology, Haringhata 741249, West Bengal, India\\
$^c$Department of Mathematics, Jadavpur University, Kolkata 700032, West Bengal, India}
\date{Received: date / Accepted: date}

\maketitle

H~I~G~H~L~I~G~H~T
\vspace{0.5cm}
\hrule
\vspace{0.5cm}
$\bullet$ We study anisotropic strange star in $f(R,T)$ gravity under Krori-Barua spacetime.

$\bullet$ The set of solutions provides non-singular as well as stable stellar model.

$\bullet$ Using observed values for mass and radius, we calculate different physical parameters.

$\bullet$ The coupling constant $\chi$, between matter and geometry, is the key factor in $f(R,T)$ gravity.
\vspace{0.5cm}
\hrule

\begin{abstract}
In the present work, we study about highly dense compact stars which are made of quarks,
specially strange quarks, adopting the Krori-Barua (KB)~\cite{Krori1975} metric in the framework of
$f(R,T)$ gravity. The equation of state (EOS) of a strange star can be represented by the MIT bag model as
$p_r(r)=\frac{1}{3}[\rho(r)-4B_g]$ where $B_g$ is the bag constant, arises due to the quark pressure.
Main motive behind our study is to find out singularity free and physically
acceptable solutions for different features of strange stars. Here we also investigate
the effect of alternative gravity in the formation of strange stars. We find that
our model is consistent with various energy conditions and also satisfies Herrera's cracking
condition, TOV equation, static stability criteria of Harrison-Zel$'$dovich-Novikov etc.
The value of the adiabatic indices as well as the EOS parameters re-establish the acceptability
of our model. Here in detail we have studied specifically three different strange star candidates, viz.
$PSRJ~1614~2230, Vela~X-1$ and $Cen~X-3$. As a whole, present model fulfils all the
criteria for stability. Another fascinating point we have discussed is the
value of the bag constant which lies in the range $(40-45)$~MeV/fm$^{3}$. This is
quite smaller than the predicted range, i.e., $(55-75)$~MeV/fm$^{3}$
~\cite{Farhi1984,Alcock1986}. The presence of the constant ($\chi$),
arises due to the coupling between matter and geometry, is responsible behind this
reduction in $B_g$ value. For $\chi=0$, we get the higher value
for $B_g$ as the above mentioned predicted range.
\end{abstract}


\end{frontmatter}

\section{Introduction}
At the final stage of a gravitationally collapsed star, i.e., when all the
thermonuclear fuels get exhausted it turns into a neutron star.  In the year
1932, the particle neutron was discovered by Chadwick and soon after this
discovery the actuality of neutron star was predicted. Later, this concept was
strongly confirmed through the observational evidences from pulsars~\cite{Hewish1968}.

The density of neutron stars is enormously high that can bend the spacetime fabric,
 mostly dominated by neutrons along with a negligible fraction of
electrons and protons. Neutron star is very small in expanse, radius of it
extends upto 11 to 15 km~\cite{Demorest2010} and mass about 1.4 to 2 solar mass
 ($M_\odot$)~\cite{Demorest2010}. So the baryon
density in a neutron star is extremely high and more than the nuclear saturation
density $n_S \approx 0.16 fm^{-3}$ (where nucleons start striking one another)
depending on the explicit object. But the calculations show that the density
at the centre of any massive neutron star becomes four or more times greater than the
nuclear saturation density. This emphasizes regarding the high probability
of a neutron star to deconfine into a quark-gluon plasma.

Due to enormous density, the energy level of the hyperon at the Fermi-surface
becomes higher than its rest mass. This phenomenon indicates that these particles could
deconfine into strange quarks which are the most stable quarks and form
strange stars. Strange stars are basically consist of up $(u)$, down
$(d)$ and strange $(s)$ quarks but mostly dominated by strange $(s)$ quarks.
There are examples of potential candidates for strange stars available in the literature such as
$SAX~J~1808.4-3658 (SS2), SAX~J~1808.4-3658 (SS1)$, X-ray binaries at low mass as $4U~1820-30$
etc.

In this connection we represent the equation of state (EOS) for a strange star as
\begin{equation}
p_r(r)=\frac{1}{3}\left[\rho(r) -4B_g\right], \label{eq1}
\end{equation}
which is known as the bag model (as proposed by MIT group) where $B_g$ is the
bag constant. There are many literatures
~\cite{Brilenkov2013,Maharaj2014,Paulucci2014,Panda2015,Isayev2015,Abbas2015,Arbanil2016,Lugones2017}
available based on MIT bag model EOS for studying the strange stars
and their stellar structures on the background of Einstein's general theory of relativity (GR).
However, in the present work we are curious to deal with this EOS in the modified gravity.

In 1915 Einstein introduced GR, which has been continuously proving it's necessity
to resolve huge number of unrevealed mysteries of the universe. However, recently on the basis of some observational
facts~\cite{Riess1998,Perlmutter1999,Bernardis2000,Hanany2000,Peebles2000,Padmanabhan2003,Clifton2012},
Einstein's GR is facing a fundamental challenge as it is not sufficient enough to explain some of the physical phenomena. Astrophysical
observations prove the accelerated expansion of the universe via the SNeIa measurement~\cite{Perlmutter1999}, later supported by many other observations~\cite{Jain2003,Tegmark2004,Eisentein2005,Spergel2007}. Dark energy, a mysterious energy component is often introduced as blameworthy for accelerating universe.

However, origin of this accelerating energy as well as accelerating universe mechanism is still
to recognize, due to inconsistency in quantum gravity theory. Though, `cosmological
constant' is the simplest and most natural solution to explain cosmic acceleration but
problems arise duo to fine-tuning and huge dissimilarities from theory to
observations~\cite{Copeland2006,Nojiri2007,Tsujikawa2010}.
Purposeful progress has been done in dark energy model redesigning the
Einstein-Hilbert action in geometry part. This phenomenological approach
is recognized as Modified gravity, consistent with observational data
~\cite{Nojiri2003,Carroll2004,Starobinsky2007,Bamba2008,Setare2011,Jamil2011,Hussain2012}
which could have been adopted to explain the unsolved issues of the universe.

Several group of astrophysicists time to time propound several theories
on modified gravity, few of them like $f(R,T)$ gravity, $f(T)$ gravity and $f(R)$
gravity acquire greater attention than the rests. There are lot of works
available in literature under the background of alternative gravity, such as $f(R,T)$, $f(T)$, $f(R)$ etc
~\cite{Abbas2014,Abbas2015a,Abbas2015b,Abbas2015c,Zubair2016}.
Harko et al.~\cite{Harko2011} extended the $f(R)$ gravity theories
by incorporating the trace $(T)$ of energy-momentum tensor along with Ricci
scalar $R$, namely as $f(R,T)$ gravity. These alternative theories of
gravity have gone through several tests in various field of astrophysics
as well as cosmology
~\cite{Moraes2014,Moraes2015,Moraes2016a,Moraes2016b,Moraes2016c,Moraes2016d,Correa2016}
even in thermodynamics~\cite{Sharif2012,Momeni2016a}.
Various astrophysical compact objects and also theories for gravitational waves
have been studied under the background of these theories
~\cite{Shamir2015,Noureen2015,Moraes2016e,Alves2016,Das2017}.

In the present article we have attempted to explore the strange stars with
spherically symmetric and anisotropic matter distribution in $f(R,T)$
 gravity incorporating the {\it ansatz} provided by Krori and Barua (KB)
~\cite{Krori1975}. The KB spacetime involves in a well behaved metric function
and completely free from any singularity - this is the main reason behind
the choice of the KB metric in the present manuscript to obtain a
physically valid solution to the Einstein field equations. Literature survey
shows that this ansatz has been used by several authors to explore
different features of compact stars either in general relativity or in alternative gravity
\cite{Rahaman2012,Kalam2012,Hossein2012,Kalam2013,Bhar2015a,Bhar2015b,Bhar2015c,Abbas2015a,Abbas2015b,Abbas2015c,Abbas2015d,Momeni2016,Deb2018a}. We notice that Rahaman et al.~\cite{Rahaman2012} studied the strange star
with KB spacetime under the framework of Einstein's GR whereas
Deb et al.~\cite{Deb2018a} have investigated the same under $f(R,T)$ gravity
without admitting the KB spacetime. Motivated from these works we have
combined the two ideas and have studied strange stars in $f(R,T)$ gravity
admitting KB metric potentials.

So the scheme of the work is as follows: the basic
mathematics of $f(R,T)$ gravity and their solutions for strange stars have been
provided in Secs. 2 and 3. In Sec. 4 we have discussed the related
boundary conditions and determined the unknown constants whereas
model parameters have been found out in Sec. 5. Stability as well as
different features of our proposed model have been studied in Sec. 6.
Finally we have made some conclusions on the present strange stellar model in Sec. 7.

\section{Basic mathematical formalism of $f(R,T)$ gravity}
According to $f(R,T)$ theory \cite{Harko2011}, we can describe action as
\begin{equation}
S=\int d^4x \pounds_m\sqrt{-g} +\frac{1}{16\pi}\int d^4x f(R,T)\sqrt{-g}.\label{eq2}
\end{equation}

In the above expression, $f(R,T)$ is an arbitrary function of the Ricci scalar $R$ and the
trace of the energy momentum tensor, $T$. On the other hand, $g$ is the determinant of
the metric $g_{\mu\nu}$ and $\pounds_m$ being the matter Lagrangian which predicts
the possibility of a non-minimal coupling between matter and geometry. Here, $\pounds_m=-P $
represents the total pressure and in geometrical units we assume $c = G = 1$.

To derive the field equations in $f(R,T)$ gravity, we can vary the
action (\ref{eq2}) w.r.t. the metric tensor $g_{\mu\nu}$ as
\begin{eqnarray}
R_{\mu\nu}f_R(R,T)&-&\frac{1}{2}g_{\mu\nu}f(R,T)+f_R(R,T)(g_{\mu\nu}\square-\nabla_\mu\nabla_\nu) \nonumber\\
 = 8\pi\mathcal{T}_{\mu\nu}&-&\mathcal{T}_{\mu\nu}f_T(R,T)-\Theta_{\mu\nu}f_T(R,T),\label{eq3}
\end{eqnarray}
where $f_T(R,T)=\frac{\partial f(R,T)}{\partial T}$ , $f_R(R,T)=\frac{\partial f(R,T)}{\partial R}$, $\square\equiv\frac{\partial_\mu(\sqrt{-g}g^{\mu\nu}\partial_\nu)}{\sqrt{-g}}$, $R_{\mu\nu}$
denotes the Ricci tensor, $\nabla_\mu$ is the covariant derivative w.r.t.
the symmetry connected to $g_{\mu\nu}$, $\Theta_{\mu\nu}=g^{\alpha\beta}\frac{\delta \mathcal{T}_{\alpha\beta}}{\delta g^{\mu\nu}}$ and $\mathcal{T}_{\mu\nu}=g_{\mu\nu}\pounds_m-2\frac{\partial\pounds_m}{\partial g^{\mu\nu}}$ is the stress-energy tensor.

From Eq. (\ref{eq3}), the covariant divergence is as follows~\cite{Barrientos2014}
\begin{eqnarray}
&\qquad\hspace{-1.8cm}\nabla^\mu \mathcal{T}_{\mu\nu}=\frac{f_T(R,T)}{8\pi-f_T(R,T)}[(\Theta_{\mu\nu}+\mathcal{T}_{\mu\nu})\nabla^\mu\ln f_T(R,T)-\frac{1}{2}g_{\mu\nu}\nabla^\mu T+\nabla^\mu\Theta_{\mu\nu}].\label{eq4}
\end{eqnarray}

Eq. (\ref{eq4}) says that in $f(R,T)$ theory of gravity, energy-momentum tensor is
not conserved where as it remains conserved in general relativity.

For a perfect anisotropic fluid we have the energy-momentum tensor in the following form
\begin{equation}
\mathcal{T}_{\mu\nu}=(\rho+p_t)u_\mu u_\nu-p_t g_{\mu\nu}+(p_r-p_t)v_\mu v_\nu, \label{eq5}
\end{equation}
with $u^\mu\nabla_\nu u_\mu=0$ and $u^{\mu}u_{\mu}= 1$. Here $\rho(r)$,
$p_r(r)$, $p_t(r)$, $u_\mu$ and $v_\mu$ stand for the energy density, radial pressure,
tangential pressure, four-velocity and radial four-vector respectively for a static fluid source. Besides
these, we have another condition $\Theta_{\mu\nu}=-2\mathcal{T}_{\mu\nu}-pg_{\mu\nu}$.

Following the  proposal by Harko et al.~\cite{Harko2011}, we can assume
the form of $f(R,T)$  as
\begin{equation}
f(R,T)=R+2\chi T.  \label{eq6}
\end{equation}

Here $\chi$ is coupling constant due to modified gravity. This form of $f(R,T)$
gravity is astronomically useful to obtain several cosmological solutions~\cite{Reddy2013b,Moraes2014b,Singh2015,Moraes2015a,Moraes2015b,Kumar2015,Shamir2015}.

By substituting the above form of $f(R,T)$ in Eq. (\ref{eq3}), we get
\begin{equation}
G_{\mu\nu}=8\pi \mathcal{T}_{\mu\nu}+\chi T g_{\mu\nu}+2\chi(\mathcal{T}_{\mu\nu}+pg_{\mu\nu}).\label{eq7}
\end{equation}

Here $G_{\mu\nu}$ denotes the Einstein tensor. We can regain the results of general
relativity just by putting $\chi=0$ in the above Eq. (\ref{eq7}).

Now combining Eqs. (\ref{eq4}) and (\ref{eq6}), eventually we have
\begin{equation}
(8\pi+2\chi)\nabla^{\mu}\mathcal{T}_{\mu\nu}=-2\chi\left[\nabla^{\mu}(pg_{\mu\nu})+\frac{1}{2}g_{\mu\nu}\nabla^{\mu}T\right].\label{eq8}
\end{equation}

Curiously, setting $\chi=0$ in Eq. (\ref{eq8}), we can verify that energy-momentum
tensor remains invariant as in general relativity.

\section{Solution of Einstein's field equations}
The line element for a static, spherically symmetric spacetime of
a strange star can be described as given below
\begin{equation}
ds^2 =e^{\nu(r)}dt^2-e^{\lambda(r)}dr^2-r^2(d\theta^2+sin^2\theta d\phi^2),\label{eq9}
\end{equation}
where $\lambda(r)$ and $\nu(r)$ are metric potentials. Here we have chosen
$\lambda(r)=Ar^2$ and $\nu(r)=Br^2+C$ as KB type~\cite{Krori1975}.
It is to note that $A$, $B$ and $C$ are random constants which can be evaluated depending on several
physical requirements.
In this proposed model, for the energy-momentum tensor, non-zero components
are given by
\begin{eqnarray}
\mathcal{T}_0^0 &=& \rho(r), \label{eq10}\\
\mathcal{T}^1_1 &=& -p_r(r), \label{eq11}\\
\mathcal{T}^2_2 &=& \mathcal{T}^3_3=-p_t(r).\label{eq12}
\end{eqnarray}

For a static uncharged fluid source, the Einstein field equations (EFE) can be represented as
\begin{eqnarray}
&\qquad\hspace{-5cm}\frac{e^{-\lambda}}{r^2}(-1+e^\lambda+r\lambda')=8\pi\rho+2\chi\left[2\rho-\frac{p_r+2p_t}{3}\right] =8\pi\rho^{eff},\label{eq13} \\
&\qquad\hspace{-5.0cm}\frac{e^{-\lambda}}{r^2}(1-e^{\lambda}+r\nu')=8\pi p_r-2\chi\left[\rho-\frac{4p_r+2p_t}{3}\right]=8\pi p_r^{eff},\label{eq14} \\
&\qquad\hspace{-2.1cm}\frac{e^{-\lambda}}{4r}\left[2(\nu'-\lambda')+(2\nu''+\nu'^2-\lambda'\nu')r\right]=
8\pi p_t-2\chi\left[\rho-\frac{p_r}{3}-\frac{5p_t}{3}\right]=8\pi p_t^{eff}  .\label{eq15}
\end{eqnarray}
Here $'\prime'$ represents the differentiation of the respective parameters w.r.t.
the radial parameter $r$ and
\begin{eqnarray}
&\qquad\hspace{-1.1cm}\rho^{eff}=\rho+\frac{\chi}{4\pi}\left[2\rho-\frac{p_r+2p_t}{3}\right],   \label{eq15a} \\
&\qquad\hspace{-1.01cm}p_r^{eff}=p_r-\frac{\chi}{4\pi}\left[\rho-\frac{4p_r+2p_t}{3}\right],\label{eq15b} \\
&\qquad\hspace{-1.25cm}p_t^{eff}=p_t-\frac{\chi}{4\pi}\left[\rho-\frac{p_r+5p_t}{3}\right]. \label{eq15c}
\end{eqnarray}

Using metric potentials $\lambda(r)=Ar^2$, $\nu(r)=Br^2+C$ and their first
order derivatives in Eqs. (\ref{eq1}), (\ref{eq13})-(\ref{eq15}), we can solve
\begin{eqnarray}
&\qquad\hspace{-9.5cm}\rho(r)=\frac{3}{4} {e^{-Ar^2}} \frac{(A+B)}{\chi+4\pi}+B_g, \label{eq16}\\
&\qquad\hspace{-9.8cm}p_r(r)=\frac {e^{-Ar^2}}{4}\frac{(A+B)}{\chi+4\pi}-B_g, \label{eq17}\\
&\qquad\hspace{-1.7cm}p_t(r)=\frac{e^{-Ar^2}\left[\left(3Br^2(B-A)+A+10B \right)\chi-12\pi \left(ABr^2-B^2r^2+A-2B \right)\right]}{2(5\chi+12\pi)(\chi+4\pi)}+32\chi(\frac{\chi}{4}+\pi)B_g.\label{eq18}
\end{eqnarray}

Using Eqs. (\ref{eq16})-(\ref{eq18}), we can calculate the values
of $\rho^{eff}, p_r^{eff}, p_t^{eff}$ as Eqs. (\ref{eq15a})-(\ref{eq15c}).

The anisotropic stress can be expressed as
\begin{eqnarray}
&\qquad\hspace{-2cm}\Delta=[p_t^{eff}- p_r^{eff}]=\frac{3}{4\pi(12\pi+5\chi)}\left[-\frac{1}{2}(ABr^2-B^2r^2+\frac{A}{2}-\frac{5B}{2})\chi e^{-Ar^2}+ \right.\nonumber\\
&\qquad\hspace{-0.1cm}\left.\pi(B-A)(2Br^2+3)e^{-Ar^2}+B_g(16\pi^2+3\chi^2+16\pi\chi)\right].\label{eq19}
\end{eqnarray}

\section{Boundary conditions}

\subsection{Interior spacetime}
From Eq.~(\ref{eq16}) we can evaluate the effective density function at the centre
\begin{eqnarray}
&\qquad\hspace{-8cm}\rho_0^{eff} =\rho^{eff}(r=0) \nonumber \\
&\qquad\hspace{-1.3cm}=\frac{3}{4(12\pi+5\chi)}\left[\frac{1}{4\pi}\chi(9A+5B)+3(A+B)+\frac{1}{\pi}(4\pi+3\chi)B_g(\chi+4\pi)\right].\label{eq20}
\end{eqnarray}

According to anisotropic condition, the radial pressure balances the tangential
pressure at the center ($r=0$), i.e.
\begin{eqnarray}
&\qquad\hspace{-0.2cm}p_r^{eff}(r=0)=p_t^{eff}(r=0).\label{eq21}
\end{eqnarray}

Using expression for $B_g$, we can compute the coupling constant ($\chi$) due
to modified gravity for different strange stars. Though solving Eq. (\ref{eq21}),
we get three values of $\chi$ for each strange star, one value is positive and small
whereas the other two are negative and high. Here, we have shown and explained all
the characteristics for positive as well as negative $\chi$ values for all the strange stars
under consideration. However, it is observed that negative $\chi$ satisfies neither
Herrera's cracking condition~\cite{Herrera1992} nor any other stability criteria. So, we
have restricted ourselves for discussions in details for positive $\chi$ value only.

\subsection{Exterior spacetime}
In the exterior region, as there is no
mass, coupling constant $\chi$ due to the modified gravity becomes zero. All the
components of the energy momentum tensor $\mathcal{T}_{\mu\nu}=(\rho+p_t)u_\mu u_\nu-p_t
g_{\mu\nu}+(p_r-p_t)v_\mu v_\nu$ are also zero
at exterior which leads to Schwarzschild solution for static exterior, as follows
\begin{eqnarray}
&\qquad\hspace{-.5cm}ds^2=\left(1-\frac{2M}{r}\right)dt^2-\left(1-\frac{2M}{r}\right)^{-1}dr^2-r^2(d\theta^2+\sin^2\theta d\phi^2),\label{eq21a}
\end{eqnarray}
where $M$ is the total mass of the stellar body. At the boundary $r=\Re$
(where $\Re$ is the radius) the metric coefficients $g_{tt}$, $g_{rr}$ and
$\frac{\partial g_{tt}}{\partial r}$ are continuous between the exterior
and interior region.

Hence, by comparing Eqs. (\ref{eq9}) and (\ref{eq21a}), we get
\begin{eqnarray}
&&g_{tt} = 1-\frac{2M}{\Re} = e^{B\Re^2+C},\label{eq22} \\
&&g_{rr} = 1-\frac{2M}{\Re} = e^{-A\Re^2}, \label{eq23} \\
&&\frac{\partial g_{tt}}{\partial r} = \frac{M}{\Re^2} = B\Re e^{B\Re^2+C}. \label{eq24}
\end{eqnarray}

Solving Eqs. (\ref{eq22})-(\ref{eq24}) we can express the constants $A$,
$B$ and $C$ in terms of $\Re$ and $M$ as follows
\begin{eqnarray}
&\qquad\hspace{-1cm}A=-\frac{1}{\Re^2}\ln\left(1-\frac{2M}{\Re}\right),\label{eq25} \\
&\qquad\hspace{-2.2cm}B=\frac{M}{\Re^2(\Re-2M)},\label{eq26} \\
&\qquad\hspace{-0.2cm}C=\ln\left(1-\frac{2M}{\Re}\right)-\left(\frac{M}{\Re-2M}\right).\label{eq27}
\end{eqnarray}

The radial pressure disappears at the boundary $(r=\Re)$, i.e.,
\begin{eqnarray}
&\qquad\hspace{-0.9cm}p^{eff}_r(r=\Re)=\frac{1}{4(12\pi+5\chi)}\left[\frac{(-4AB\Re^2+4B^2\Re^2-7A+5B)\chi{e^{-A\Re^2}}}{4\pi} \right.\nonumber \\
&\qquad\hspace{-2cm}\left.+3(A+B)e^{-A\Re^2}-3(4\pi+3\chi)(\chi+4\pi)B_g\right]=0.\label{eq28}
\end{eqnarray}

Putting the values of $A$ and $B$ we get
\begin{eqnarray}
&\qquad\hspace{-0.8cm}B_g=\frac{1}{\Re^3(\chi+4\pi)(4\pi+3\chi)}\left[\alpha\left(2M(\pi-\frac{5\chi}{12})-(\pi-\frac{7\chi}{12})\Re\right)+\right.\nonumber\\
&\qquad\hspace{-1.2cm}\left.\frac{2M}{(-\Re+2M)}\left((\frac{\chi}{4}+\pi)M-\frac{\Re}{2}(\pi+\frac{5\chi}{12})\right)\right],\label{eq29}
\end{eqnarray}
where $\alpha=\ln(1-\frac{2M}{\Re})$.

\section{Physical parameters of the proposed model}
Here, in the present study, we consider three different strange star candidates, viz. $PSR~J~1614~2230$, $Vela~X-1$
and $Cen~X-3$ along with their mass and radius~\cite{Deb2016} as shown in Table 1.

\begin{table}[!htp]
  \centering
      \caption{Mass and radius of different strange star candidates}\resizebox{\columnwidth}{!}{
      \begin{tabular}{ccccc}
\hline  Case      & Stars             & Mass $(M_\odot)$    & Radius ($\Re$ in km)    & $\frac{M}{\Re}$       \\
\hline  I         & PSR J 1614 2230   & 1.97                & 10.977                 & 0.1795                \\
\hline  II        & Vela X-1          & 1.77                & 10.654                 & 0.1661                \\
\hline  III       & Cen X-3           & 1.49                & 10.136                 & 0.1471                \\
\hline \label{Table1}
\end{tabular}}
\end{table}

From the values of $M$ and $\Re$ in Table 1, and also by using Eqs. (\ref{eq21}), (\ref{eq25}),
(\ref{eq26}) and (\ref{eq29}) we can evaluate the unknown parameters $\chi$, $A$, $B$
and $B_g$ which are represented in Table 2.\\

\begin{table}[!htp]
  \centering
      \caption{Determination of model parameters $A$, $B$, $B_g$ and $\chi$ for different strange star candidates}\resizebox{\columnwidth}{!}{
      \begin{tabular}{ccccc}
\hline Case      & $A$ {\small{(${km}^{-2}$)}}        & $B$ {\small{(${km}^{-2}$)}}    &$\chi$         & $B_g$ {\small{($MeV/{fm}^3$)}}\\
\hline I         & 0.003689961987                     & 0.002323332389                 & 0.6746867583   & 45.3       \\
\hline II        & 0.003558090580                     & 0.002191967045                 & 0.8143412761   & 43       \\
\hline III       & 0.003388625404                     & 0.002026668572                 & 1.006296869    & 40       \\
\hline \label{Table2}
\end{tabular}}
\end{table}

To verify the physical acceptability of our proposed model, we can recalculate the
value for the bag constat with $\chi=0$ from Eq.~(\ref{eq29}). In case of $\chi=0$
which represents the GR, we are getting higher values for the bag constant
for all the strange stars we considered here. This discrepancy arises due to the
effect of modified gravity. These calculated higher $B_g$ values, shown in Table. 3,
are exactly in the specified range $(55-75)~ MeV/fm^3$~\cite{Farhi1984,Alcock1986}
for the bag constant for stable strange quark matter.\\

\begin{table}[!htp]
  \centering
      \caption{Comparison of $B_g$: Modified Gravity and GR}\resizebox{\columnwidth}{!}{
      \begin{tabular}{ccccc}
\hline Case      &$\chi$        & $B_g$ {\small{($MeV/{fm}^3$)}} &$\chi$    & $B_g$ {\small{($MeV/{fm}^3$)}}  \\
\hline I         & 0.6746867583 & 45.3                           & 0        & 58.2  \\
\hline II        & 0.8143412761 & 43                             & 0        & 58    \\
\hline III       & 1.006296869  & 40                             & 0        & 57.7  \\
\hline \label{Table3}
\end{tabular}}
\end{table}

\begin{table}[!htp]
  \centering
      \caption{Determination of the effective central density, surface density, radial pressure and surface
      redshift for different strange star candidates}\resizebox{\columnwidth}{!}{
      \begin{tabular}{ccccc}
\hline  Case  & $\rho^{eff}(r=0)$       & $\rho^{eff}(r=\Re)$     & $p_r^{eff}(r=0)$        & $Z_s$   \\
              & {\small{(${gm/cm}^3$)}} & {\small{(${gm/cm}^3$)}} & {\small{(${dyne/cm}^2$)}}         \\
\hline  I     & $5.959 \times 10^{14}$  & $3.52\times 10^{14}$    & $10.3\times 10^{34}$   & $0.25$   \\
\hline  II    & $5.746 \times 10^{14}$  & $3.517\times 10^{14}$   & $9.51\times 10^{34}$   & $0.224$  \\
\hline  III   & $5.4725\times 10^{14}$  & $3.52\times 10^{14}$    & $8.53\times 10^{34}$   & $0.20$   \\
\hline \label{Table4}
\end{tabular}}
\end{table}

\section{Physical features of the proposed model}

\subsection{Density and pressure}
Solving Eqs. (\ref{eq13})-(\ref{eq15}) and the MIT bag model in Eq. (\ref{eq1}), we
can measure the effective density as given in Eq. (\ref{eq15a}) and density at the centre
as shown in Eq. (\ref{eq20}). In Fig.$1$, variation of the effective density w.r.t. $\frac{r}{\Re}$
has been shown graphically, where one can observe that at $r\rightarrow0$, the
density is very high and attains it's maximum value. For example, in case of $PSR~J~1614-2230$, $\rho^{eff}_0=
5.959\times 10^{14}\ gm/{cm}^3$ and $\rho^{eff}_{\Re}= 3.52\times 10^{14}\ gm/{cm}^3$,
whereas for $Vela X-1$, $\rho^{eff}_0= 5.746\times 10^{14}gm/{cm}^3$ and
$\rho^{eff}_{\Re}= 3.517\times 10^{14}\ gm/{cm}^3$. Though, the effective density gradually falls
towards the surface, however it demands very high matter density throughout the stellar system.

\begin{figure}[!htbp]
\centering
\includegraphics[width=6cm]{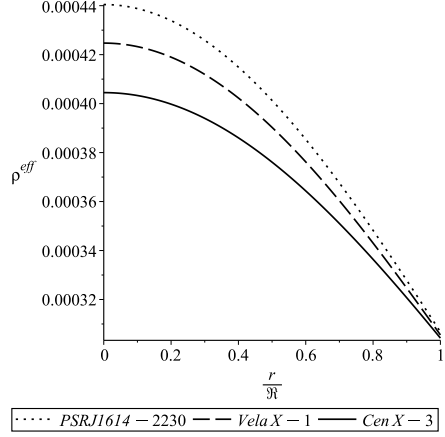}
\caption{Variation of the effective density w.r.t. the fractional radial coordinate $r/\Re$ for different
strange star candidates where $\Re$ is the radius of the corresponding star shown in Table~\ref{Table1}.
This explanation will be followed for all the other figures.}\label{pres.}
\end{figure}

\begin{figure}[!htbp]
\centering
\includegraphics[width=6cm]{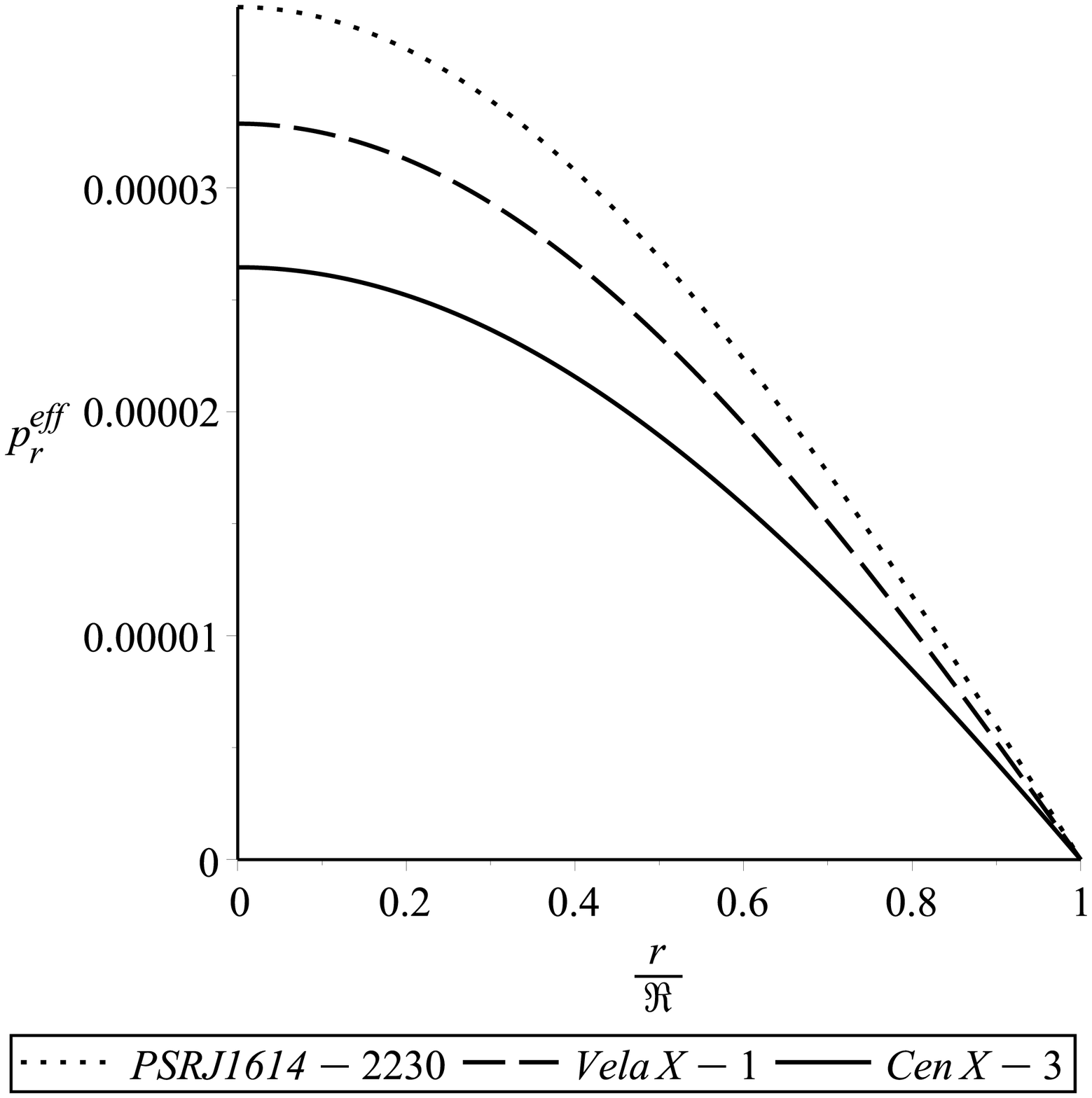}
\includegraphics[width=6cm]{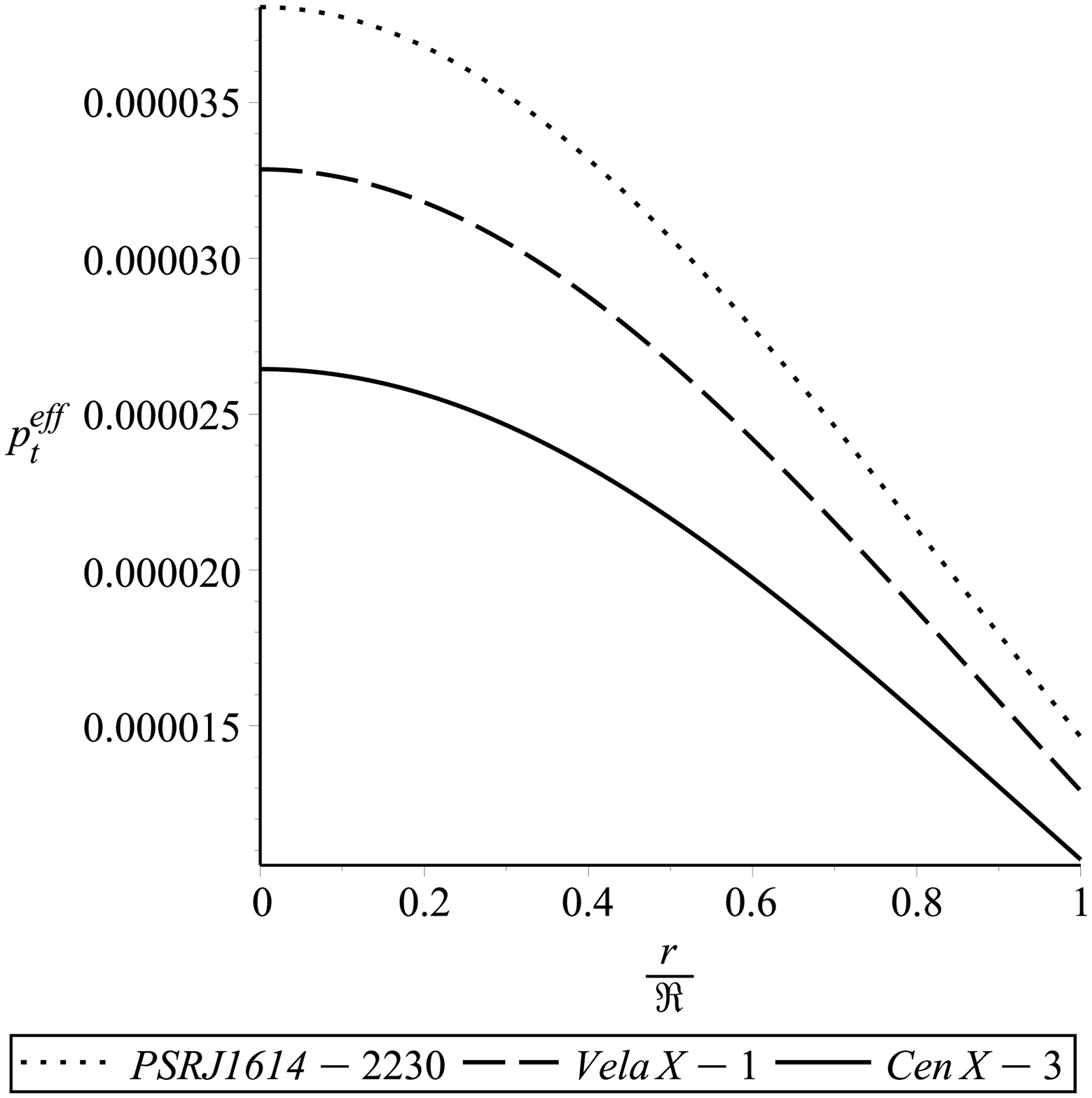}
\caption{Variation of the effective radial pressure (left panel), effective tangential pressure
 (right panel) w.r.t. the fractional radial coordinate $r/\Re$ for different strange star candidates.}\label{press.}
\end{figure}

The effective radial pressure ($p^{eff}_r$) and the effective tangential pressure
($p^{eff}_t$), related to Eqs. (\ref{eq17}) and (\ref{eq18}), have been shown graphically in Fig.~\ref{press.}.
These figures clearly indicate that both $p^{eff}_r$ and $p^{eff}_t$ are maximum at the origin
($r=0$) as in the case of the density profile and decrease gradually towards the surface.
The effective radial pressure vanishes at the surface ($r={\Re}$) from where we can
verify the size of our investigated stars.

\begin{figure}[!htbp]
\centering
\includegraphics[width=6cm]{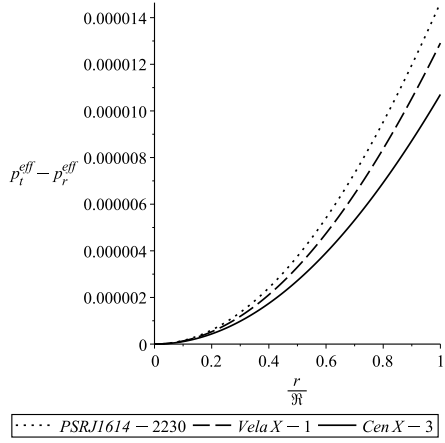}
\caption{Variation of the anisotropic stress w.r.t. the fractional radial coordinate $r/\Re$ for different
strange star candidates.}\label{aniso.}
\end{figure}

In this model, anisotropy ($\Delta$) can be defined as Eq.~(\ref{eq19}).
Hossein et al.~\cite{Hossein2012} explained that for $\Delta>0$,  i.e., $p_t>p_r$, direction of the
anisotropy will be outward and for $\Delta<0$, i.e., $p_t<p_r$, anisotropy will be inward.
The variation of the anisotropic stress $\Delta$ ($=p^{eff}_t-p^{eff}_r$)
has been displayed in Fig.~\ref{aniso.}. This figure clearly shows the zero anisotropy at the centre
of the star and then nonlinearly increasing nature throughout the stellar body. Finally,
anisotropy reaches it's maximum value at the surface which has been demanded as the inherent
nature by Deb et al.~\cite{Deb2017} for ultra-dense star. Following Gokhroo and Mehra~\cite{Gokhroo1994}
we can exhibit that the positive anisotropy leads our model to achieve a stable configuration.

\subsection{Conservation equation}
The conservation equation, i.e., Tolman-Oppenheimer-Volkoff (TOV) equation has been checked to study about the stability of our model. For an anisotropic star under equilibrium, form of generalized TOV equation can be expressed as
\begin{eqnarray}
&\qquad\hspace{-1.25cm}-\frac{(\rho+p_r)\nu'}{2}-\frac{dp_r}{dr}+
\frac{\chi}{(8\pi+2\chi)}\left[\frac{d}{dr}(\frac{p_r+2p_t}{3})-\frac{d\rho}{dr}\right]+\frac{2}{r}(p_t-p_r)=0. \label{eq30}
\end{eqnarray}

In Eq. (\ref{eq30}), there are four different forces, viz., the gravitational $(F_g)$,
hydrostatic $(F_h)$, anisotropic stress $(F_a)$ and force due to modified gravity $(F_{mg})$ so that
\begin{equation}
F_g+ F_h+F_a+F_{mg}=0, \label{eq31}
\end{equation}
with

\begin{eqnarray}
&\qquad\hspace{-7.2cm}F_g=-\frac{Bre^{-Ar^2}(A+B)}{\chi+4\pi},    \\   \label{eq32}
&\qquad\hspace{-7.2cm}F_h=\frac{1}{2}\frac{Are^{-Ar^2}(A+B)}{\chi+4\pi}, \\  \label{eq33}
&\qquad\hspace{-1cm}F_a=\frac{e^{-Ar^2}}{2r(12\pi+5\chi)(\chi+4\pi)}\left[3(2B^2r^2-2ABr^2-A+5B)\chi\right.\nonumber\\
&\qquad\hspace{-0.2cm}\left.-24\pi(A-B)(Br^2+3/2)\right]+\frac{6B_g(3\chi+4\pi)}{r(12\pi+3\chi)}, \\ \label{eq34}
&\qquad\hspace{-1.6cm}F_{mg}=\frac{(-\chi)}{(4\pi+\chi)}\left[\frac{2}{3}\frac{Are^{-Ar^2}(A+B)}{\chi+4\pi}+\frac{Br(B-A)e^{-Ar^2}}{5\chi+12\pi}-\right.\nonumber\\
&\qquad\hspace{-0.9cm}\left.\frac{\left(3Br^2(B-A)(\chi+2\pi)+\chi(A+10B)-6\pi(A-2B)\right)Are^{-Ar^2}}{3(\chi+4\pi)(5\chi+12\pi)}\right]. \\  \nonumber
\end{eqnarray}\label{eq35}

In Fig. 4, we have plotted the variation of different forces w.r.t. the radial
parameter $(r)$ for different strange stars. The plots clearly indicate that
the combined effect of anisotropic force and hydrostatic force balances
the effect of gravitational force and modified gravity force, and our
considered strange star model achieves an stable equilibrium condition under
TOV stability criteria.

\begin{figure}
\centering
\includegraphics[width=4.4cm]{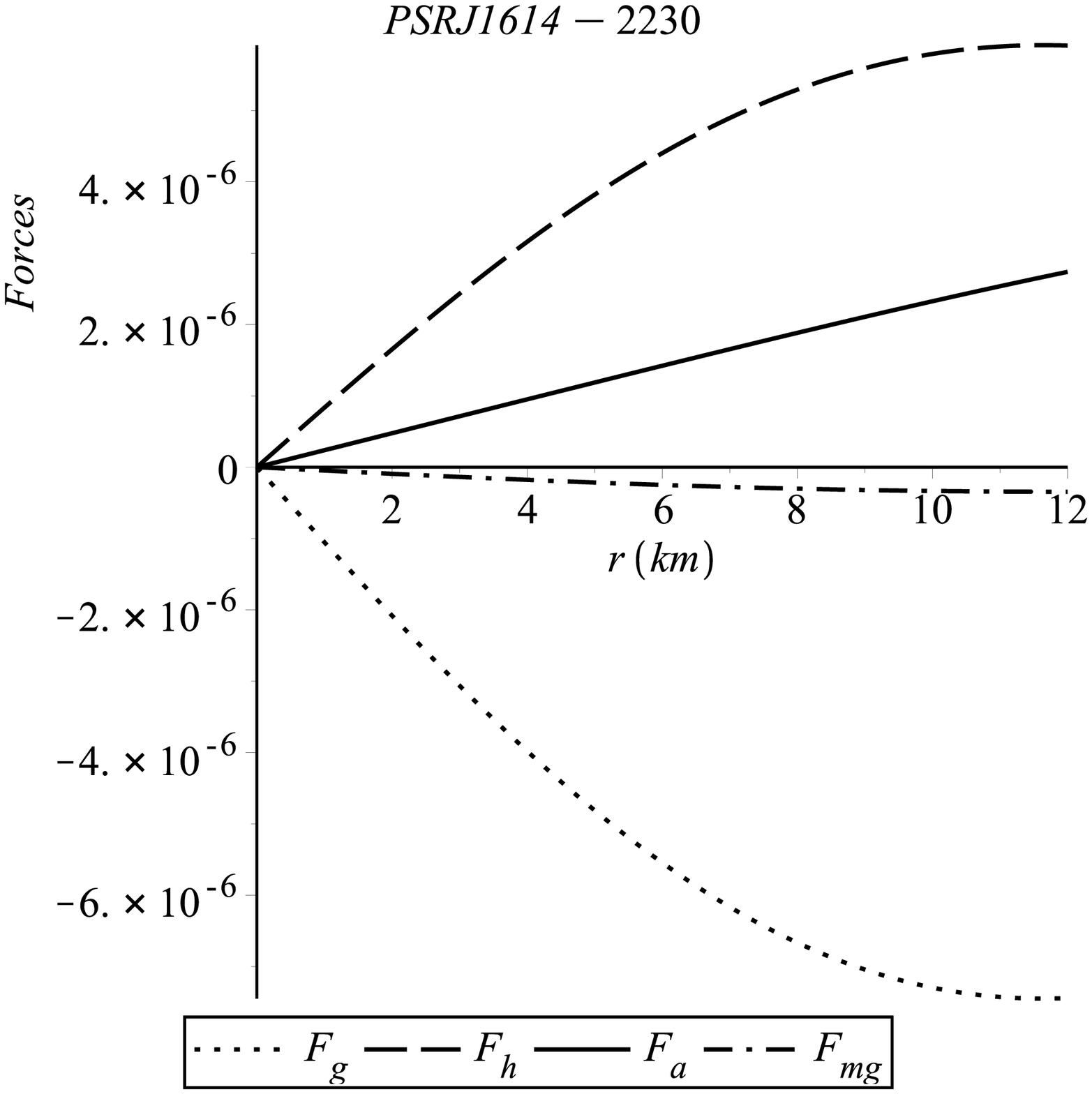}
\includegraphics[width=4.4cm]{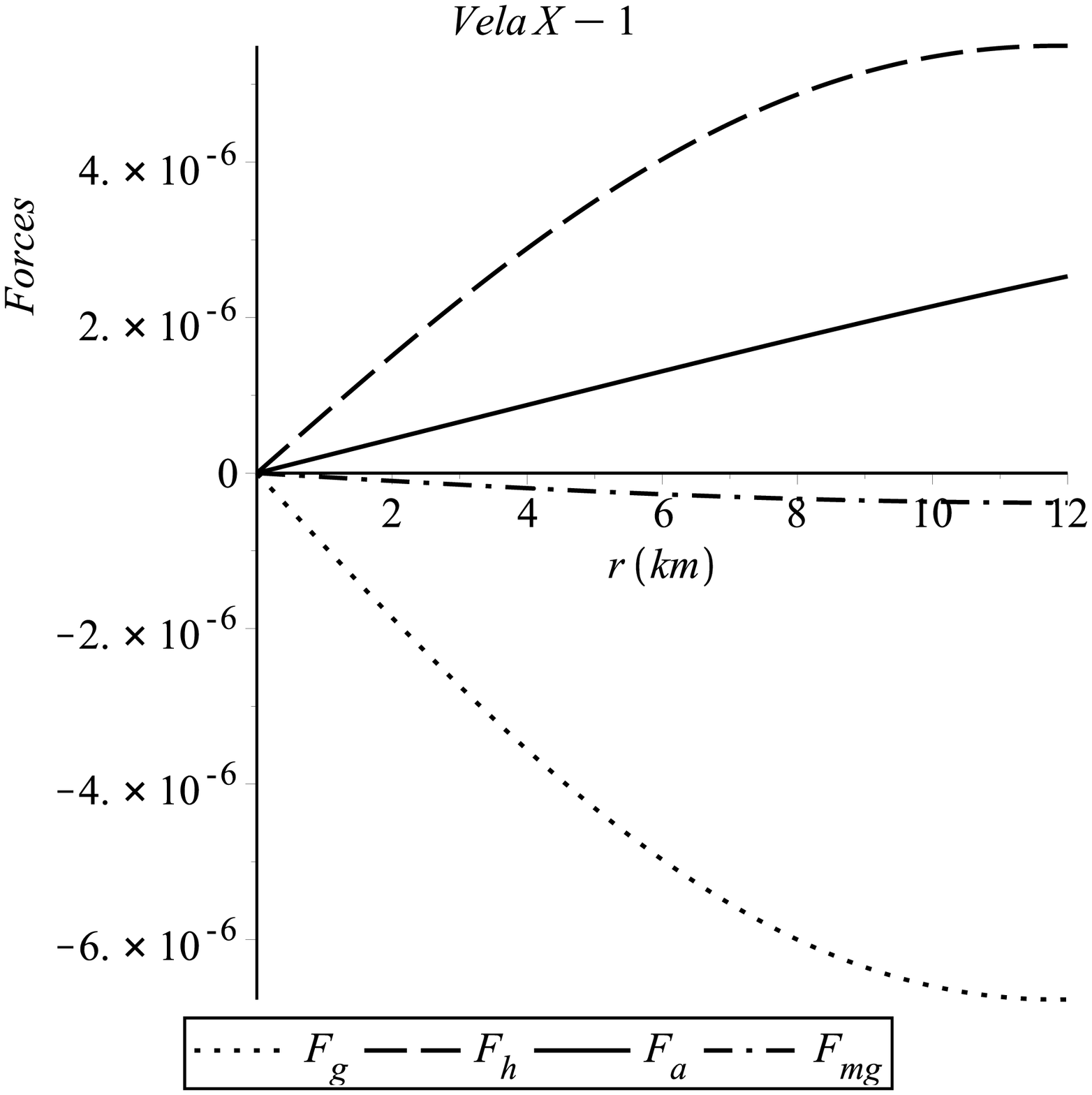}
\includegraphics[width=4.4cm]{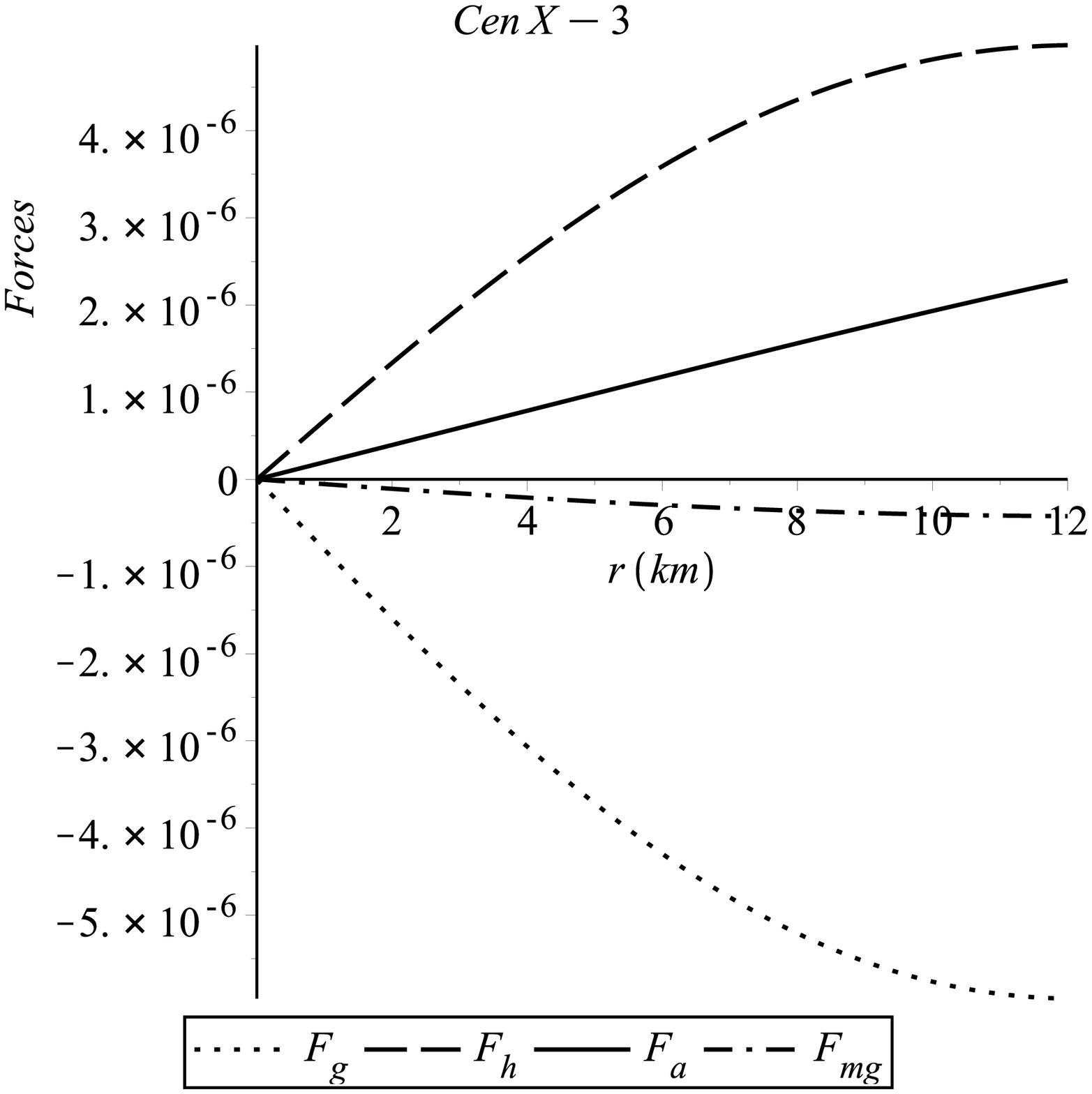}
\caption{Variation of different forces w.r.t. the fractional radial coordinate $r/\Re$ for different
strange star candidates.}\label{pot.}
\end{figure}

\subsection{Energy conditions}
From general relativity, the energy-momentum tensor $\mathcal{T}_{\mu\nu}$
describes the distribution of momentum, mass and stress due to the presence
of matter as well as any non-gravitational fields. The Einstein field equations,
however, not directly concern about the admissible non-gravitational fields
or state of matter in the spacetime model. Basically in GR the energy conditions
permit different non-gravitational fields and all states of matter and
also justify the physically acceptable solutions.

For a fluid sphere, composed of anisotropic strange matter, some inequality conditions like Null
Energy Conditions (NEC), Strong Energy Conditions (SEC), Weak Energy
Conditions  (WEC) and Dominating Energy Conditions (DEC) have to hold
simultaneously throughout the star, to get a stable model. These conditions are given below:
\begin{eqnarray}
NEC &:& \rho^{eff} \geq 0,      \\      \label{eq36}
WEC &:& \rho^{eff}+p_r^{eff} \geq 0 , \rho^{eff}+p_t^{eff} \geq 0, \\ \label{eq37}
SEC &:& \rho^{eff}+p_r^{eff}+2p_t^{eff} \geq 0,         \\     \label{eq38}
DEC &:& \rho^{eff}-|p_r^{eff}| \geq 0 , \rho^{eff}-|p_t^{eff}| \geq 0.  \label{eq39}
\end{eqnarray}

At the centre ($r=0$), above energy conditions give some bounds for
the model parameter $A$ and $B$. NEC, WEC, SEC, DEC demands $A>0$ ,
$B>0$, $(A-B)>0$ to be satisfied. Set of values shown in Table $2$ prove
that our proposed strange star model successfully satisfies the energy
conditions, shown in Fig. 5.

\begin{figure}[!htp]
\centering
\includegraphics[width=4.4cm]{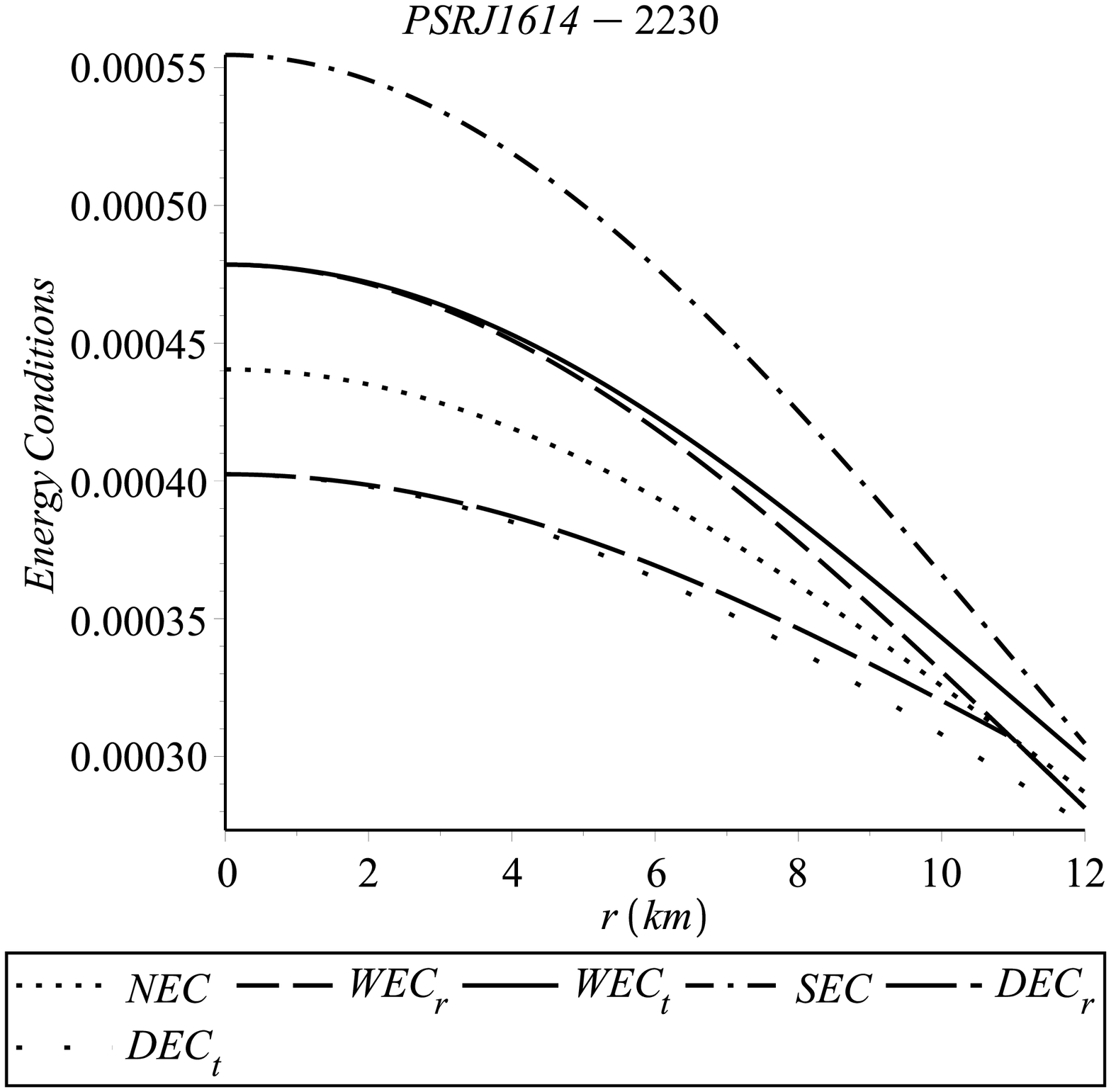}
\includegraphics[width=4.4cm]{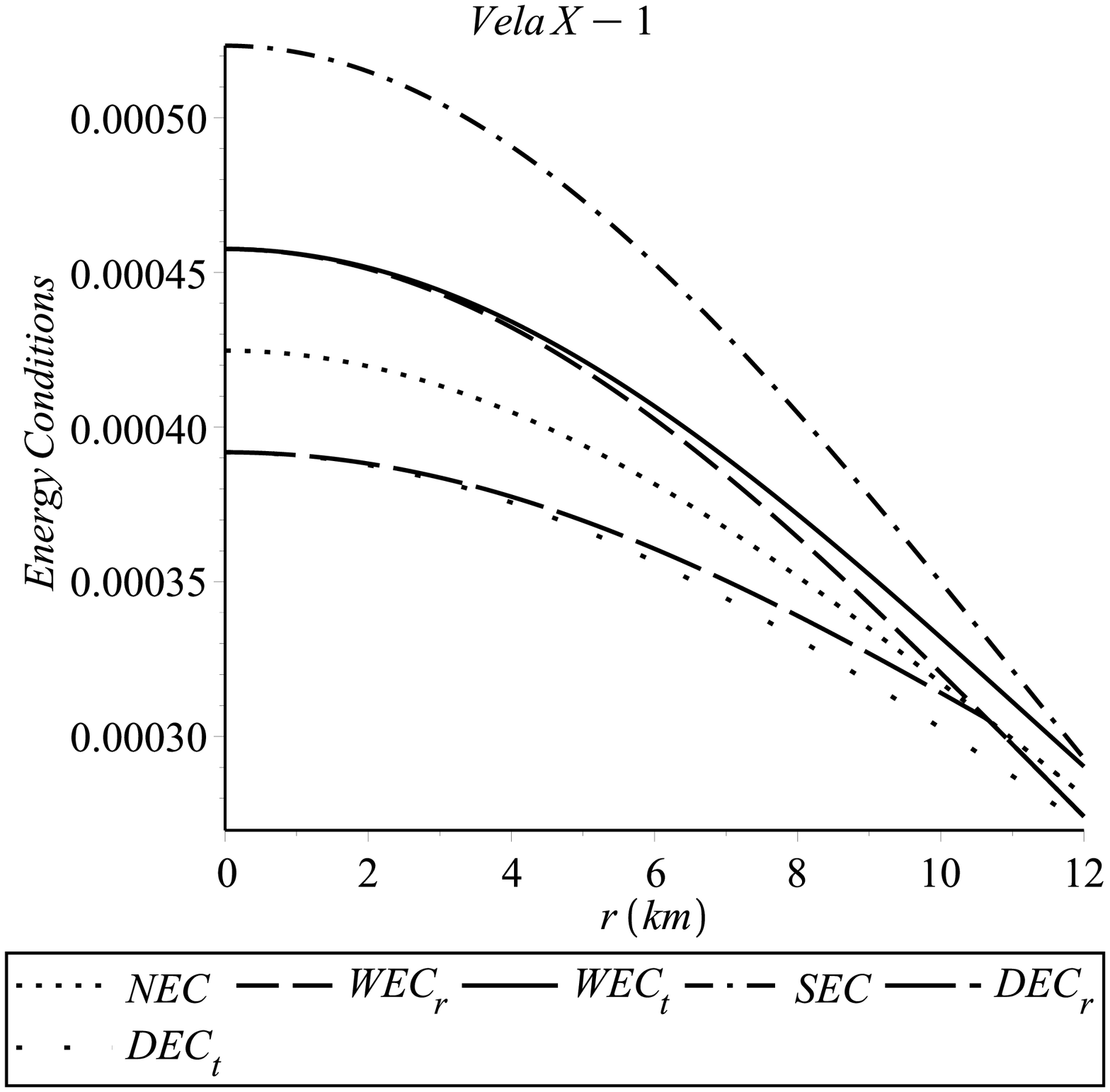}
\includegraphics[width=4.4cm]{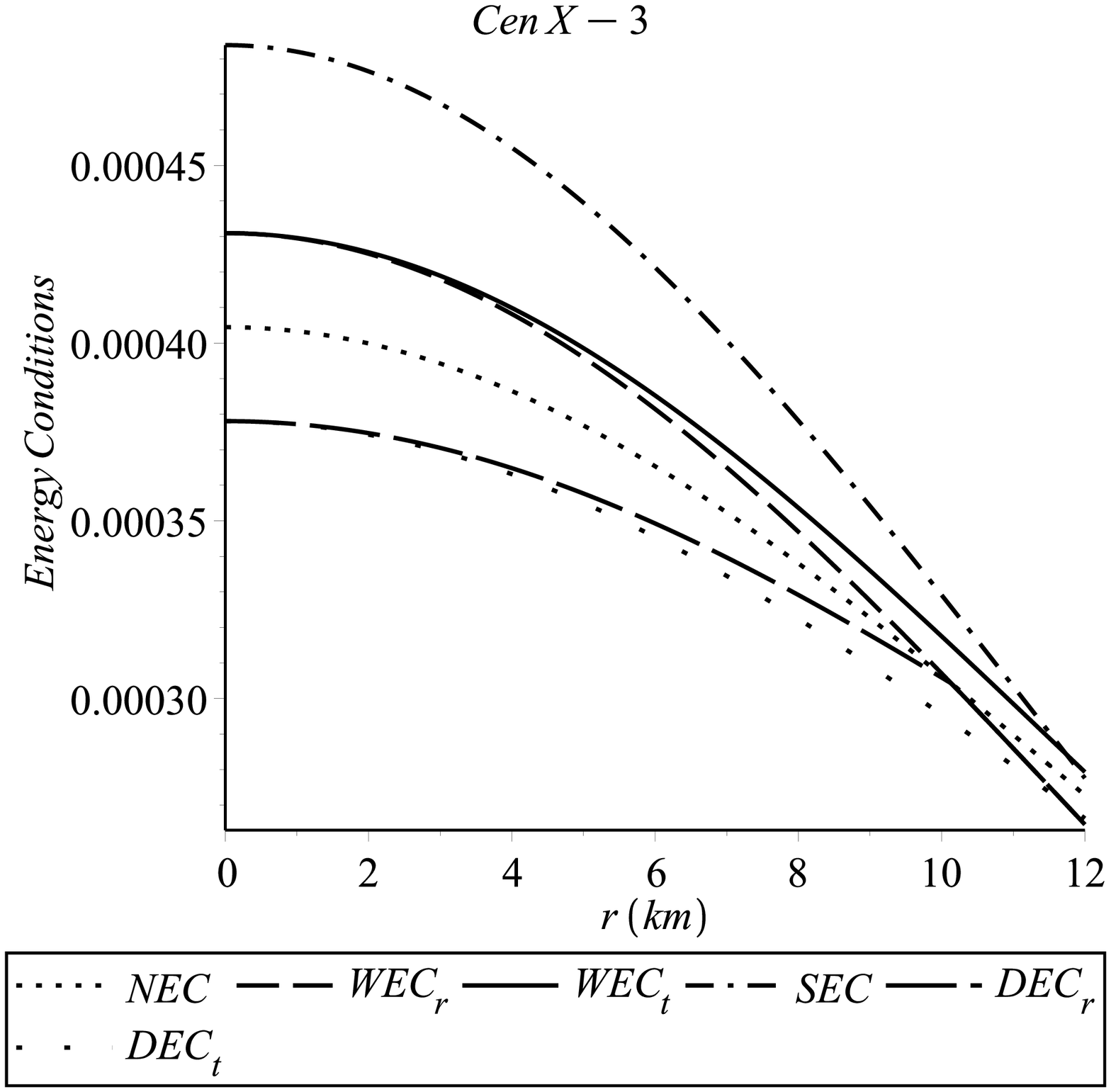}
\caption{Variation of the different energy conditions w.r.t. the fractional radial coordinate $r/\Re$ for different
strange star candidates.}\label{pres.}
\end{figure}

However, several matter distributions are there which mathematically violate SEC.
Hawking~\cite{Hawking1973} argued that SEC is not valid for any scalar field
containing a positive potential and for any cosmological inflationary process.

\subsection{Herrera's cracking condition}
The concept of cracking (breaking) appears when the equilibrium configuration of a stellar system has been
perturbed, as a result the sign of the total radial forces are different in different regions of the stellar
configuration. This cracking in the stellar system arises either from the anistropy of the fluid distribution
or due to the emission of incoherent radiation where the condition for the acceptability of anisotropic matter
distribution is $\frac{\partial p_r}{\partial \rho}<1$ and $\frac{\partial p_t}{\partial \rho}<1$, i.e., the
square of sound speed $v_{rs}^2<1$ and $v_{ts}^2<1$. With the assist of this Herrera's cracking concept~\cite{Herrera1992},
we can examine the stability of our proposed model. For a physically acceptable
fluid distribution, the causality condition demands the square of sound speed to follow
$ 0\leq v^2_{ts} \leq 1$ and $ 0\leq v^2_{rs} \leq 1$. According to Herrera
\cite{Herrera1992}, the region where radial sound speed $(v_{rs})$ dominates
the tangential sound speed $(v_{ts})$, is potentially stable. Also for
stable matter distribution, Herrera~\cite{Herrera1992} and Andr\'{e}asson
\cite{Andreasson2009} claim the condition to be imposed is $|v^2_{rs}-v^2_{ts}| \leq 1$.
This condition signifies `no cracking', i.e., the region must be potentially stable.

In our model we get the parameters as follows:
\begin{eqnarray}
&\qquad\hspace{-2.5cm}v^2_{rs}=\frac{dp^{eff}_r}{d\rho^{eff}}=\frac{4B\chi(A-B)+\chi A(-4ABr^2+B^2r^2-7A+5B)+12A\pi(A+B)}{[4B\chi(A^2r^2-ABr^2+B)+36A^2\pi +27A^2\chi+36AB\pi+11AB\chi]},  \label{eq40} \\
&\qquad\hspace{-2cm}v^2_{ts}=\frac{dp^{eff}_t}{d\rho^{eff}}=\frac{-2({A}^{2}B{r}^{2}-A{B}^{2}{r}^{2}+{A}^{2}-3AB+{B}^{2})(12\pi+5\chi)}{(4A^2B\chi r^2
-4AB^2\chi r^2+36A^2\pi+27A^2\chi+36AB\pi+11AB\chi+4B^2\chi)}.\label{eq41}
\end{eqnarray}

In GR, for any model following the MIT bag EOS, the value of the square of radial sound speed ($v_{rs}^2$)
is a constant ($1/3$). But, due to the coupling parameter ($\chi$) in the
modified gravity, $v_{rs}^2$ becomes as Eq. (\ref{eq40}), where $\chi=0$
gives back the constant result as can be achieved in GR.

\begin{figure}[!htp]
\centering
     \includegraphics[width=6cm]{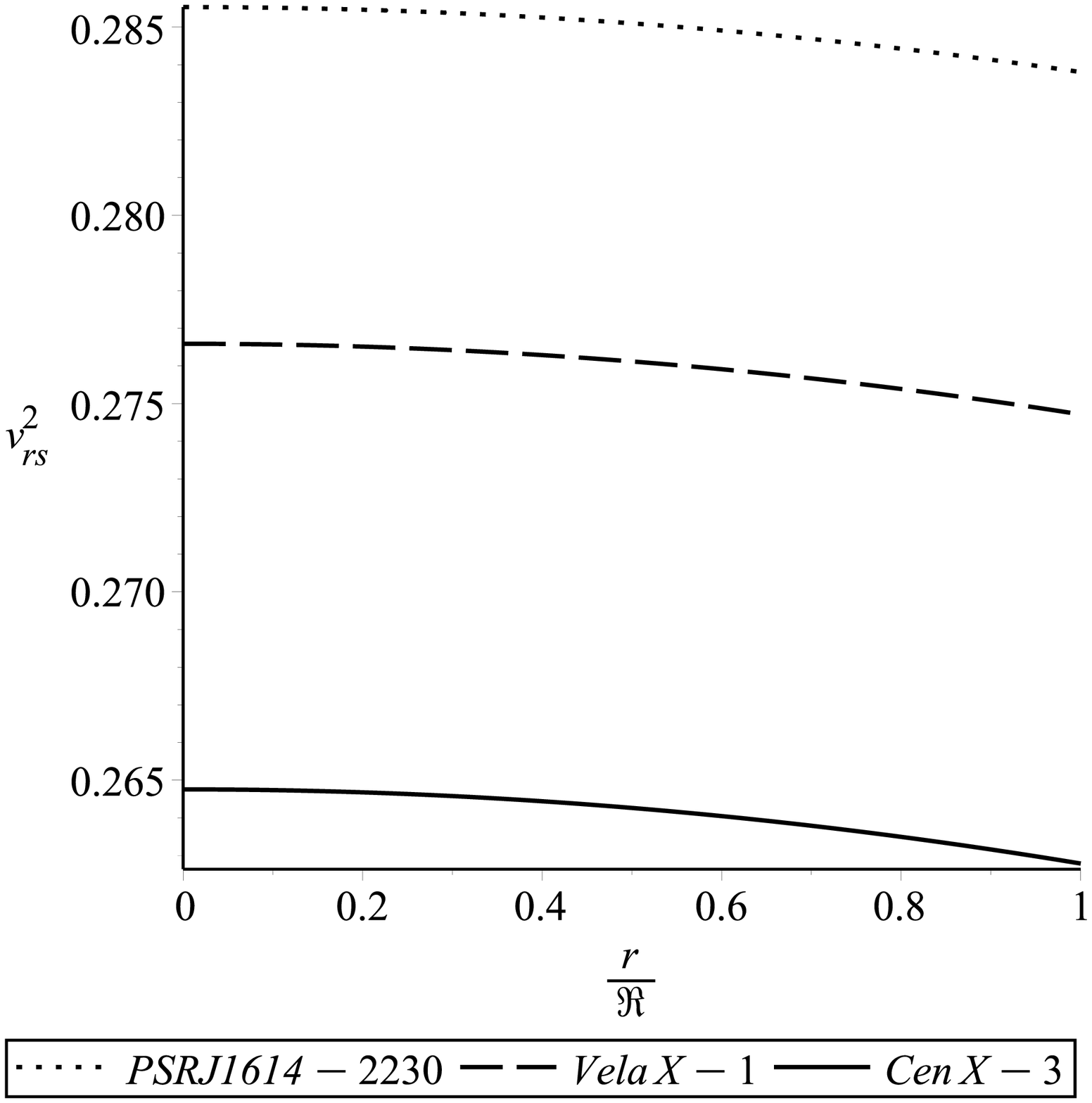}
     \includegraphics[width=6cm]{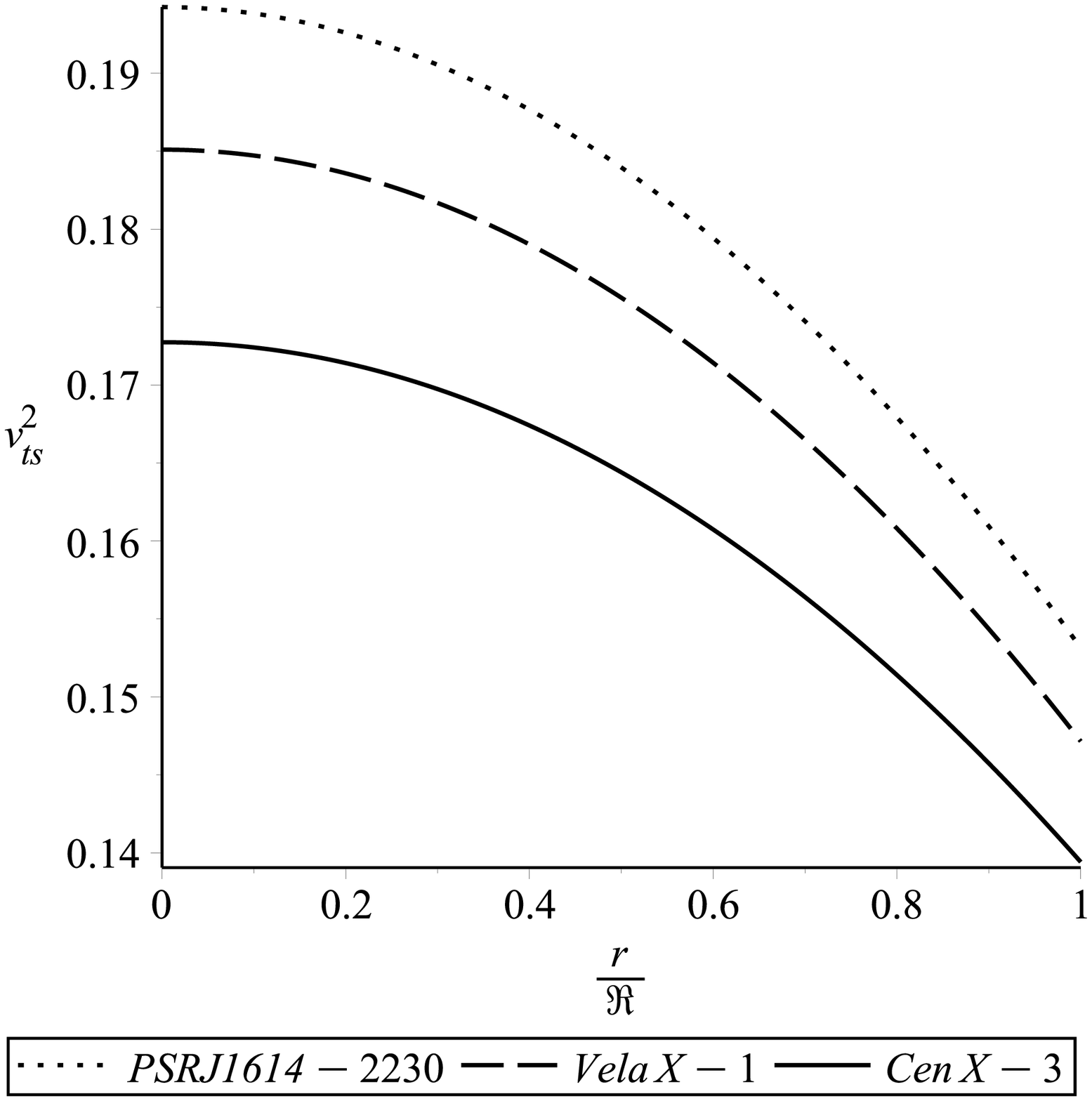}
\caption{Variation of $v^2_{rs}$ (left panel), $v^2_{ts}$ (right panel) w.r.t. the fractional
 radial coordinate $r/\Re$ for different strange star candidates.}\label{vel.}
\end{figure}

Graphical representation for causality conditions  and Herrera's cracking
condition~\cite{Herrera1992} have been shown in Fig. 6 and Fig. 7
respectively. From Fig. 6, it is clear that both $v^2_{rs}$ and
$v^2_{ts}$ are less than 1, i.e., our model is consistent with Herrera's cracking
concept and Fig. 7 also shows potential stability throughout the stars.

\begin{figure}[!htp]
\centering
     \includegraphics[width=6cm]{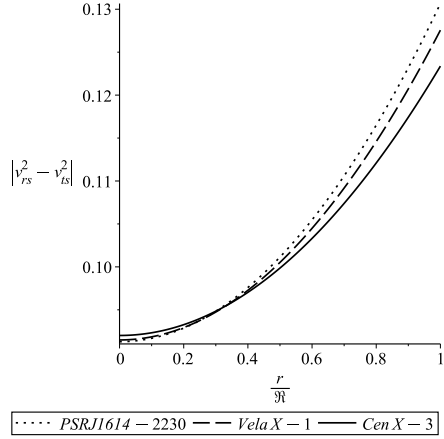}
\caption{Variation of $|v^2_{rs}-v^2_{ts}|$ w.r.t. the fractional radial coordinate $r/\Re$ for different
strange star candidates.}\label{veld.}
\end{figure}

\subsection{Effective mass and compactification factor}
For a spherically symmetric, static strange star made of perfect
anisotropic fluid, Buchdahl~\cite{Buchdahl1959} established that there
is an upper limit for the ratio of  allowed maximum mass and radius $(\Re)$,
i.e., $\frac{2M}{\Re}<\frac{8}{9}$. In our proposed model,
the gravitational effective mass takes the following form

\begin{eqnarray}
M^{eff}=\int _{0}^{\Re}4\pi r^2\rho^{eff} dr=\int _{0}^{\Re}4\pi r^2\left[\rho+\frac{\chi}{4\pi}\left(2\rho-\frac{p_r+2p_t}{3}\right)\right] dr \nonumber\\
=m+\int _{0}^{\Re}r^2\chi\left(2\rho-\frac{p_r+2p_t}{3}\right)dr, \label{eq42}
\end{eqnarray}
where $m=4\pi\int_0^{\Re}r^2\rho dr$ is mass function for the distribution
of strange quark matter and the remaining part $\int _{0}^{\Re}r^2\chi\left(2\rho-\frac{p_r+2p_t}{3}\right)dr=m_{mg}$
is another mass distribution which has been generated due to the modified gravity. For $\chi=0$ Eq. (\ref{eq42})
leads to the GR solution, i.e. $M^{eff}=m$.

\begin{figure}[!htp]
\centering
\includegraphics[width=6cm]{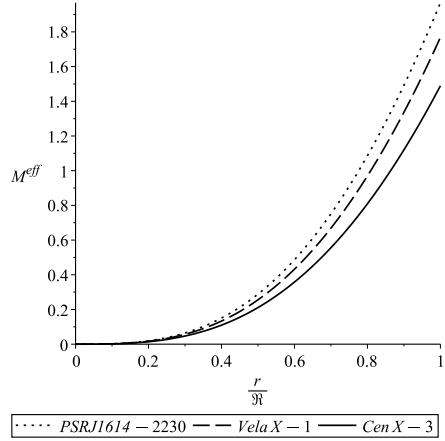}
\caption{Variation of the effective mass w.r.t. the fractional radial coordinate $r/\Re$ for different
strange star candidates.}\label{mass.}
\end{figure}

\begin{figure}[!htp]
\centering
\includegraphics[width=6cm]{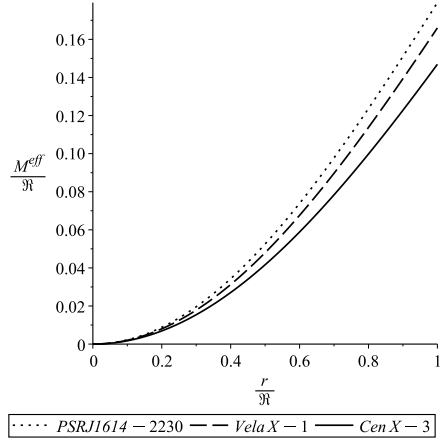}
\caption{Variation of the compactness w.r.t. the fractional radial coordinate $r/\Re$ for different
strange star candidates.}\label{cf.}
\end{figure}
The compactification factor $u(r)$ can be defined
as the ratio of the effective mass $(M^{eff})$ and radius $(\Re)$, which is given below
\begin{eqnarray}
&\qquad\hspace{-0.8cm}u(r)=\frac{M^{eff}(r)}{\Re}=\frac{1}{\Re}[m+\int _{0}^{\Re}r^2\chi\left(2\rho-\frac{p_r+2p_t}{3}\right)dr].  \label{eq43}
\end{eqnarray}

The variation of effective mass and compactification factor w.r.t.
to $\frac{r}{\Re}$ have been shown graphically in Figs. 8 and 9 respectively, where both
increase with increasing radii.

\subsection{Surface redshift}
The surface redshift is defined as
\begin{equation}
Z_s=\frac{1}{\sqrt{1-2u}}-1. \label{eq44}
\end{equation}

Barraco and Hamity~\cite{Barraco2002} showed that for an isotropic star,
$Z_s\leq2$  when the cosmological constant is absent. Later, B{\"o}hmer and Harko
~\cite{Bohmer2006} proved that surface redshift may be much higher ($Z_s\leq5$)
for an anisotropic star when the cosmological constant is present. Eventually,
this restriction get modified, and calculations show that $Z_s=5.211$ ~\cite{Ivanov2002}
is the maximum acceptable limit. In the current study, we have calculated the value
for maximum surface redshift for different strange stars and get $Z_s\leq1$
[~\ref{Table4}] in every case.

\subsection{Equation of State (EOS)}
\begin{figure}[!htp]
\centering
\includegraphics[width=5cm]{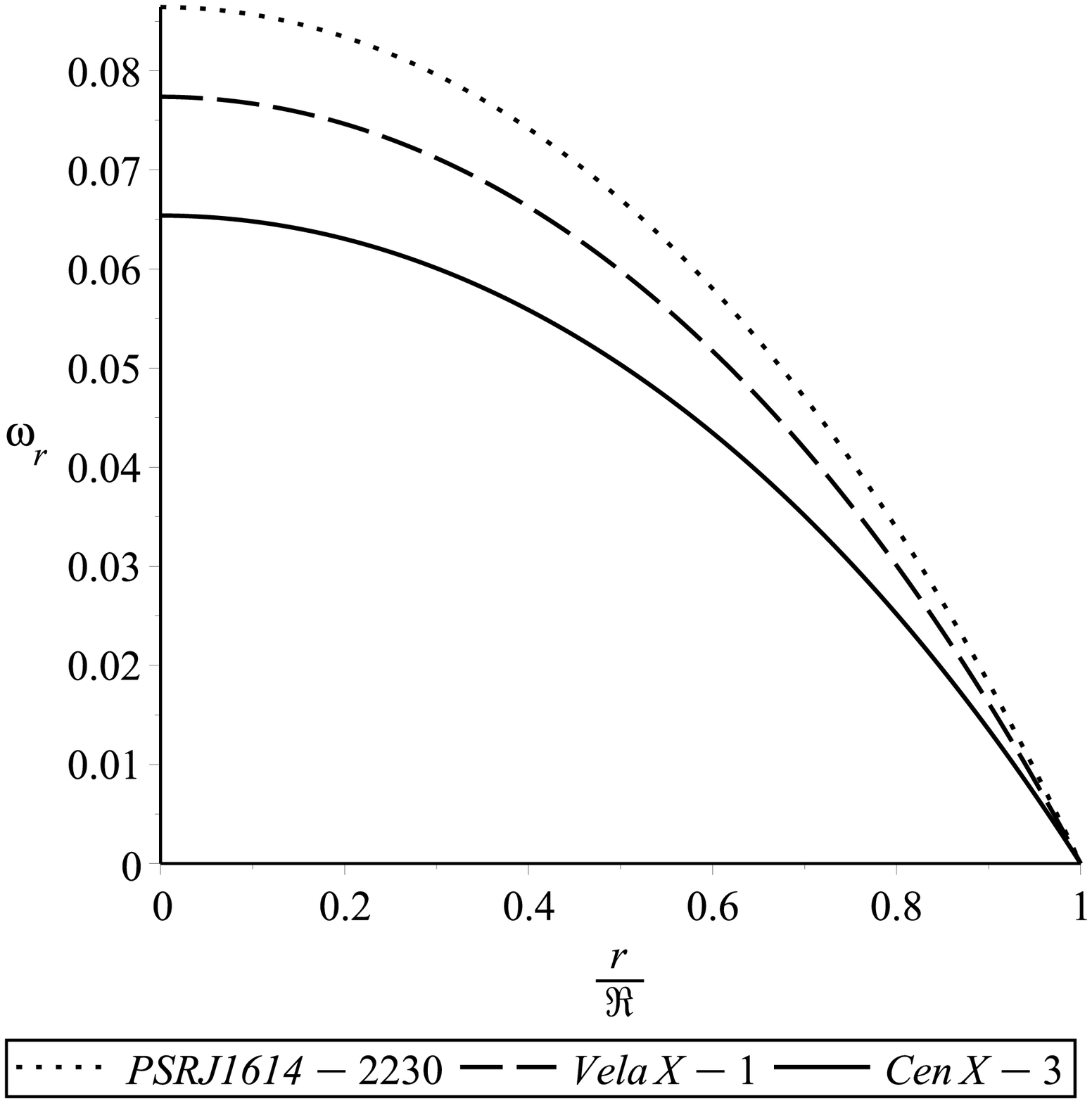}
\includegraphics[width=5cm]{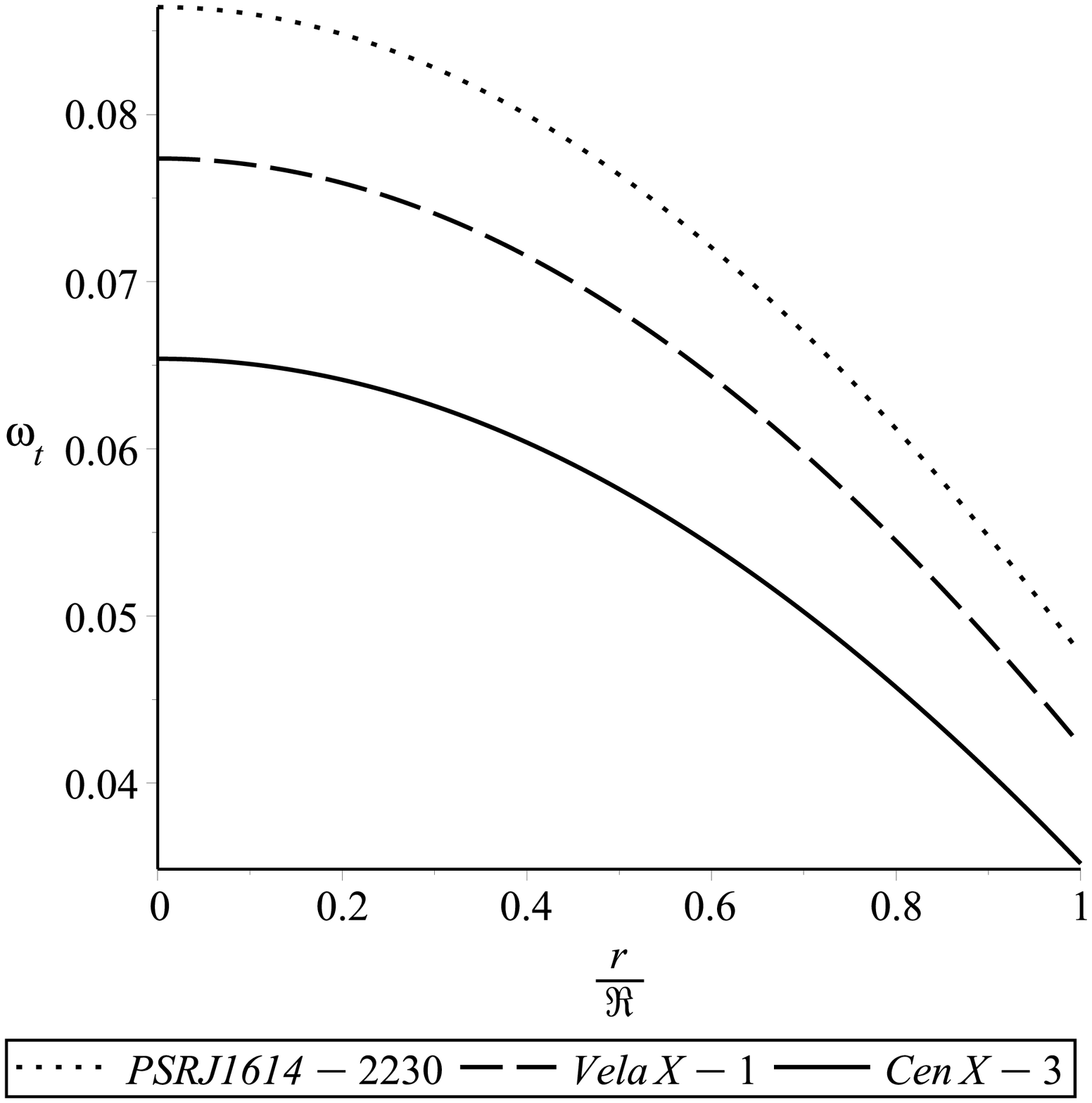}
\caption{Variation of the EOS parameter $w_r$ (left panel) and $w_t$ (right panel) w.r.t. the
fractional radial coordinate $r/\Re$ for different strange star candidates.}\label{eos.}
\end{figure}

According to our proposed model, we can represent radial ($\omega_r$) and
tangential ($\omega_t$) EOS as follows
\begin{eqnarray}
\small
&\qquad\hspace{-1cm}\omega_r=\frac{p^{eff}_r}{\rho^{eff}}=\frac{\left[\alpha\chi+12\pi(A+B)\right]{e^{-A{r}^{2}}}-12(4\pi+3\chi)B_g(\chi+4\pi)}{\left[\beta\chi+36\pi(A+B)\right]{e^{-A{r}^{
2}}}+12(4\pi+3\chi)B_g(\chi+4\pi)},\\ \label{eq45}
&\qquad\hspace{-1cm}\omega_t=\frac{p^{eff}_t}{\rho^{eff}}=\frac{-2(AB{r}^{2}-{B}^{2}{r}^{2}+A-2B){e^{-Ar^2}}(12\pi+5\chi)}{[\beta\chi+36\pi( A+B)]e^{-Ar^2}+12(4\pi+3\chi)B_g(\chi+4\pi)}, \label{eq46}
\end{eqnarray}
where $\alpha=(-4AB{r}^{2}+4{B}^{2}{r}^{2}-7A+5B)$, $\beta=(4ABr^2-4B^2r^2+27A+15B)$.
We have plotted EOS parameter $(\omega)$
w.r.t. the fractional radial coordinate $\frac{r}{\Re}$, for both EOS $\omega_r$ and
$\omega_t$, shown in Fig.~\ref{eos.}. The figures clearly show that
throughout the fluid sphere, $\omega_r$ and $\omega_t$ are
positive and they lie in $0<\omega<1$, which establish the
non-exotic nature of strange quintessence star.

\subsection{Adiabatic index}
The adiabatic index $\Gamma$ can be described as the ratio of two specific
heat~\cite{Hillebrandt1976} and for a given density profile, $\Gamma$ characterizes the stiffness of that EOS.
Chandrasekhar~\cite{Chandrasekhar1964} introduced the idea of
the dynamical stability of the stellar model against an infinitesimal radial adiabatic perturbation.
Later on this stability condition was developed and used at astrophysical level by several scientists~\cite{Bondi1964,Bardeen1966,Wald1984,Knutsen1988,Mak2013}. The stability condition demands that
the adiabatic index $\Gamma> \frac{4}{3}$. For the anisotropic relativistic sphere the radial and transverse adiabatic
index as $\Gamma_r$ and $\Gamma_t$ respectively, can be expressed as
\begin{eqnarray}
\Gamma_r &=& \frac{\rho^{eff}+p_r^{eff}}{p_r^{eff}}\left[\frac{dp_r^{eff}}{d\rho^{eff}}\right], \\ \label{eq47}
\Gamma_t &=& \frac{\rho^{eff}+p_t^{eff}}{p_t^{eff}}\left[\frac{dp_t^{eff}}{d\rho^{eff}}\right].  \label{eq48}
\end{eqnarray}

In our model, using above relations, we get the following equations
{\begin{eqnarray} \label{eq48}
&\qquad\hspace{-1.2cm}\Gamma_r=\frac{(12\pi+5\chi)\gamma\left[\left((Br^2+\frac{7}{4})A^2-(B^2r^2+\frac{9B}{4})A+B^2\right)\chi-3A\pi\gamma
\right]\phi}{[\left(\alpha\chi-3\pi\gamma\right)\phi+6\beta]\left[\left(\xi A^2+(-B^2r^2+\frac{11B}{4})A+B^2\right)\chi+9A\pi\gamma\right]}, \label{eq48}\\
&\qquad\hspace{-1.2cm}\Gamma_t=\frac{\delta\left[\left(-r^2B^2\kappa+B\left(Ar^2\kappa-\frac{7\pi}{2}-\frac{35\chi}{24}\right)
-A(\frac{\pi}{2}+\frac{17\chi}{24})\right)\phi-
\beta\right]}{\theta \phi\left[B^2\chi(1-Ar^2)+A(A\chi r^2+9\pi+\frac{11\chi}{4})B+9A^2(\pi+\frac{3\chi}{4})\right]}, \label{eq49}
\end{eqnarray}
where, $\alpha=(ABr^2-B^2r^2+\frac{7A}{4}-\frac{5B}{4})$, $\beta=\frac{B_g}{2}(4\pi+3\chi)(\chi+4\pi)$,
$\gamma=(A+B)$, $\delta=-6[(1-Ar^2)B^2+(Ar^2-3)AB+A^2]$, $\phi=e^{-Ar^2}$,
$\xi=(Br^2+\frac{27}{4})$, $\kappa=(\frac{\chi}{4}+\pi)$ and $\theta=(ABr^2-B^2r^2+A-2B)$.

\begin{figure}[!htp]
\centering
\includegraphics[width=5cm]{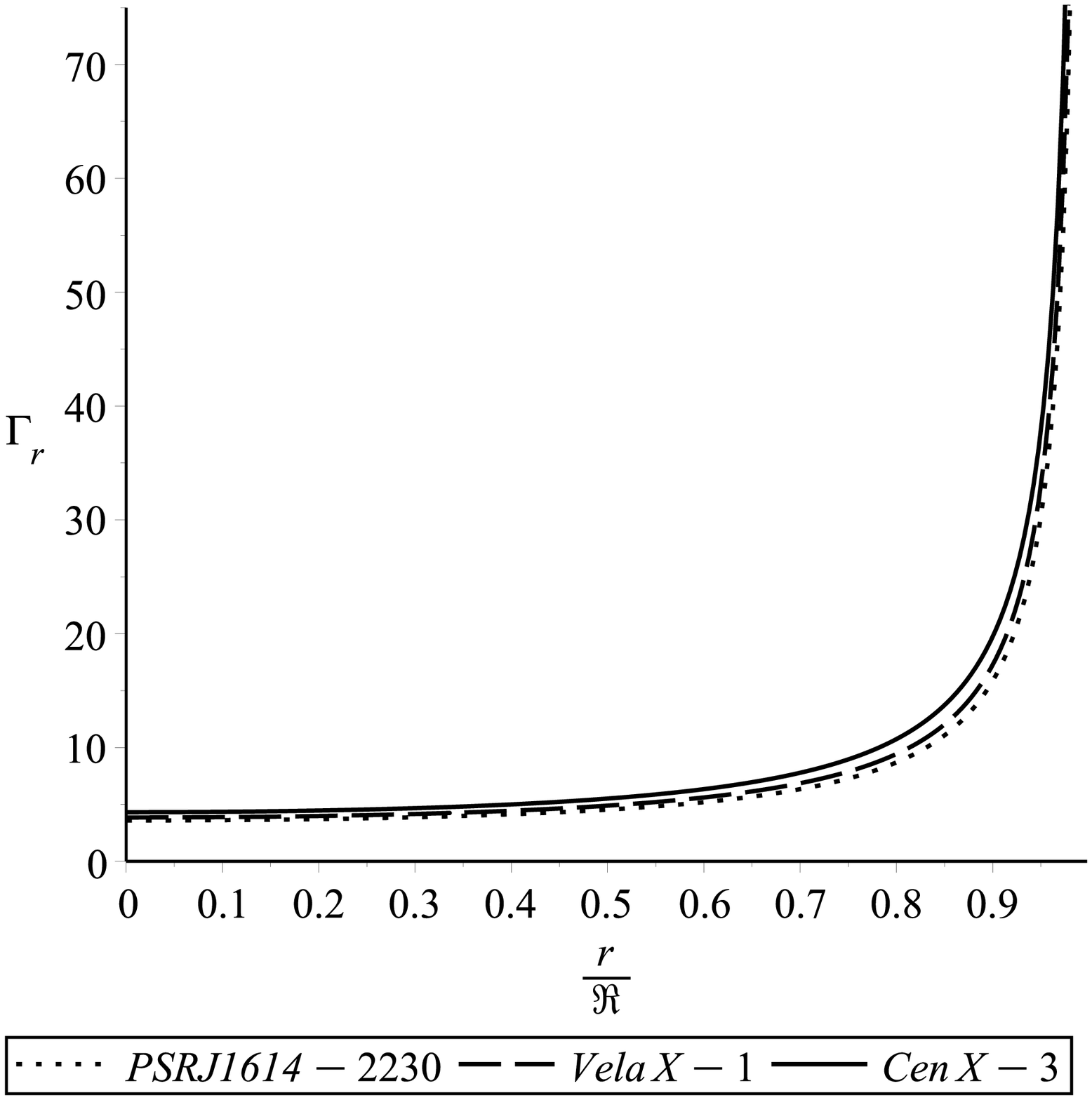}
\includegraphics[width=5cm]{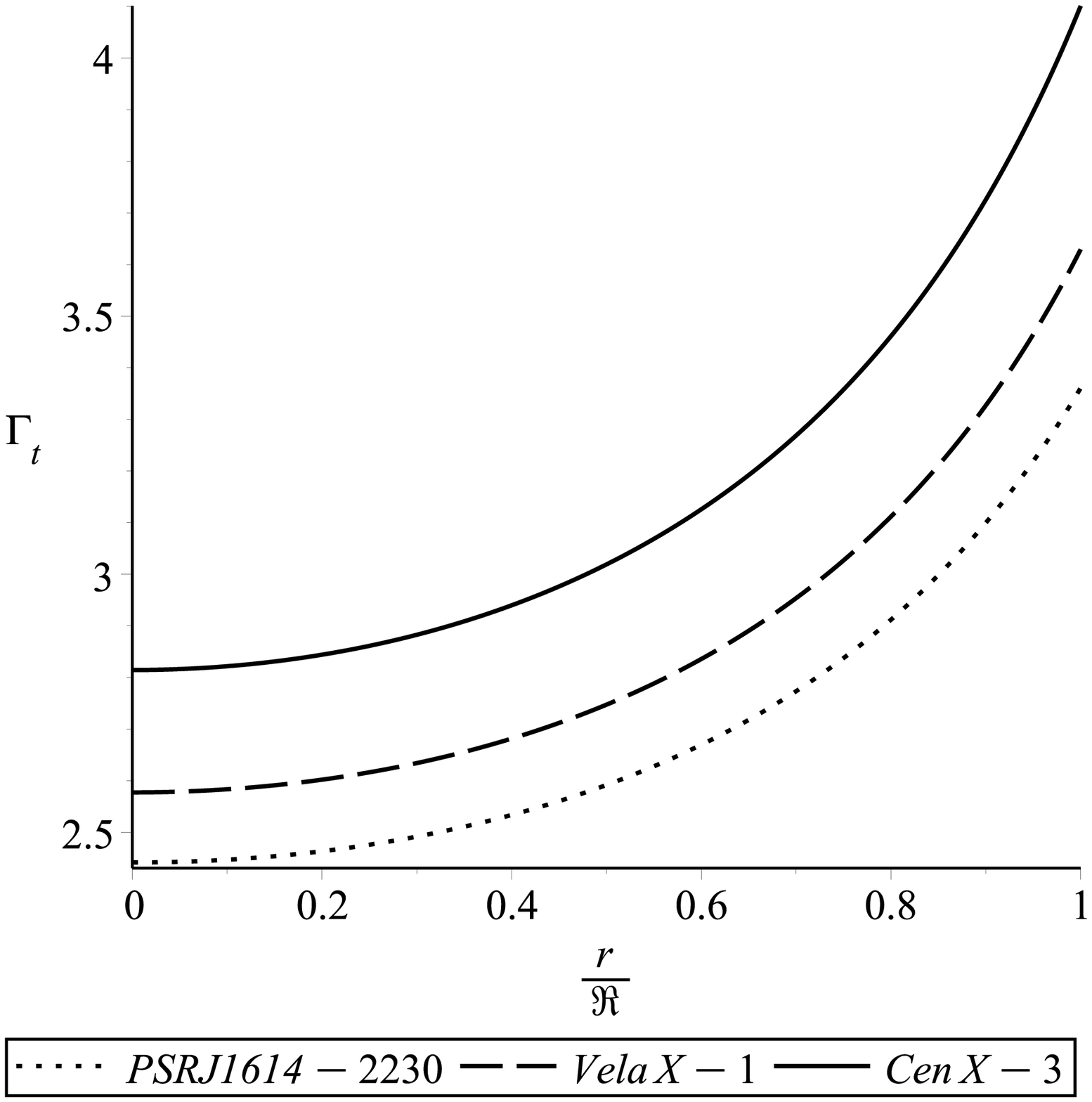}
\caption{Variation of the adiabatic index $\Gamma_{r}$ (left panel) and $\Gamma_{t}$ (right panel) w.r.t. the
 fractional radial coordinate $r/\Re$ for different strange star candidates.}\label{adia.}
\end{figure}

It can be shown analytically that both the adiabatic indices $(\Gamma_r,~\Gamma_t) > \frac{4}{3}$
 throughout the interior of the strange star and thus satisfy the stability condition. The graphical representations (Fig.~\ref{adia.}) also establish the stability of our proposed model.

\subsection{Harrison-Zel$'$dovich-Novikov static stability criteria}
In 1964, Chandrasekhar~\cite{Chandrasekhar1964a} and in 1965 Harrison et al.
~\cite{Harrison1965} calculate the eigen-frequencies for all the fundamental modes.
Later, Zel$'$dovich and Novikov~\cite{Zeldovich1971} make the calculations more
simpler following Harrison et al.~\cite{Harrison1965}. For that purpose, they assumed
that adiabatic index of a slowly deformed matter is comparable with that of a pulsating star.
From their assumption, nature of mass will be increasing w.r.t the central density
(i.e. $\frac{dM}{d\rho_c}>0$) for a stable configuration. Model will be unstable if
$\frac{dM}{d\rho_c}<0$.

In this model, mass can be expressed in terms of the central density ($\rho_c$) as follows
\begin{eqnarray}
&\qquad\hspace{-2.2cm}M(\rho_c)=\frac{R^3[16\pi(12\pi\beta-12\pi\rho_c+12\beta\chi-5\chi\rho_c)+36\chi^2\beta+36A
\pi+27A\chi]}{[32\pi R^2(12\pi\beta-12\pi\rho_c+12\beta\chi-5\chi\rho_c)+18R^2(4\beta\chi^2+4A\pi +3A\chi)-36\pi-15\chi]}.\label{eq50}
\end{eqnarray}

Differentiating Eq.~(\ref{eq50}) w.r.t. $\rho_c$ we get
\begin{eqnarray}
&\qquad\hspace{-2.2cm}\frac{dM}{d\rho_c}=\frac{48R^3\pi(12\pi+5\chi)^2}{[32\pi R^2(12\pi\beta-12\pi\rho_c+12\beta\chi-5\chi\rho_c)+18R^2(4\beta\chi^2+4A\pi +3A\chi)-36\pi-15\chi]^2}.\label{eq51}
\end{eqnarray}}

\begin{figure}[!htp]
\centering
\includegraphics[width=6cm]{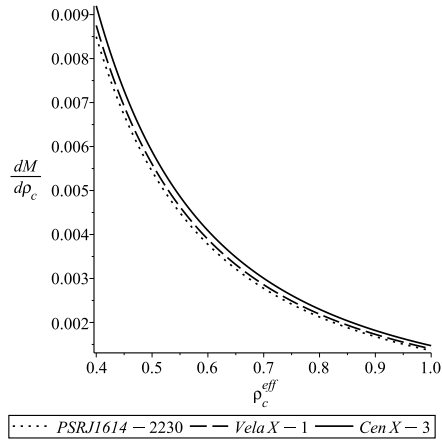}
\caption{Variation of $\frac{dM}{d\rho_c}$ w.r.t. $\rho_c$ for different strange star candidates.}\label{HZN.}
\end{figure}

From Eq.~(\ref{eq51}), it is very clear that $\frac{dM}{d\rho_c}$ is always positive
inside the star and Fig.~\ref{HZN.} also shows the positive value for $\frac{dM}{d\rho_c}$
throughout the stellar structure. So, our model fulfils the Harrison-Zel$'$dovich-Novikov condition
 and further confirms the stability~\cite{Shee2016,Bhar2017}.

\section{Discussions and conclusions}
In this paper, we have explored strange quark star in $f(R,T)$
gravity using the KB~\cite{Krori1975} metric functions, where
we assume the MIT bag model as the EOS for strange quark matter distribution. Since, the matter
distribution is assumed to be anisotropic in nature, the system is
not at all over determined due to inclusion of the EOS. With the help of
the KB metric and the MIT bag model as EOS, we have investigated here
various interesting physical features and also represented graphically,
the variation of different physical parameters w.r.t. the fractional radial coordinate $r/\Re$.

However, the specific major findings of the present investigation can be categorized as follows:

1. From our study we can predict the existence of stable strange stars in the range of lower value of $B_g$ $(40-45)~Mev/fm^3$ whereas earlier works assumed~\cite{Deb2018a,Deb2018b} or obtained~\cite{Rahaman2012,Bhar2015} higher values of $B_g$ for the construction of strange stellar models in both GR~\cite{Rahaman2012,Bhar2015} and modified theories~\cite{Deb2018a,Deb2018b}. Our investigation, therefore, clearly indicates that $f(R,T)$ gravity effectively reduces $B_g$. Here the matter-geometry coupling constant~$\chi$ plays an important role for this reduction. Setting $\chi=0$ in the results of our study, one can get the higher values of $B_g$ in the range $(55-75)~Mev/fm^3$ which is till now the proposed range~\cite{Farhi1984,Alcock1986,Weber2005} for stable strange quark matter distribution under GR. However, experimental results from RHIC and CERN-SPS, show the possibility of wide range of value for bag constant in case of density dependent bag model~\cite{Burgio2002}. In Table~\ref{Table2}, we have provided the calculated bag value from our model. At the same time, incorporating $\chi=0$ in Eq.~(\ref{eq29}) we have again calculated the bag value, which actually signifies $B_g$ in the frame of GR. In the second case, we get $B_g$ in the range $(57-60)~Mev/fm^3$, shown in Table~\ref{Table3} which strongly supports the stability criteria~\cite{Farhi1984,Alcock1986}.

2. Following the earlier works~\cite{Rahaman2012,Bhar2015,Deb2018a,Deb2018b}, we have checked the stability issue of strange stars through the studies of Herrera's cracking conditions, energy conditions, Buchdahl limit, TOV equation, EOS parameter and adiabatic index. Variations of all these parameters w.r.t. the fractional radial coordinate $r/\Re$ clearly indicate the stability of our model and physical acceptability for the construction of stable strange star under $f(R,T)$ gravity with KB spacetime. In addition to above, we have also checked
the criteria of Harrison-Zel$'$dovich-Novikov for static stability which is well satisfied. This is very crucial point to note that none of the earlier works~\cite{Rahaman2012,Bhar2015,Deb2018a,Deb2018b} satisfies all the stability criteria at a time as our study does effectively.

3. The present study can be claimed as a continuation of the earlier works~\cite{Rahaman2012,Bhar2015,Deb2018a,Deb2018b} which provides more promising results by fine tuning of various model parameters.

Now we would like to summarize all the general features of the present study as follows:

$\textbf{(i) Density and Pressure:}$
In our present investigation the effective density ($\rho^{eff}$), effective radial pressure ($p^{eff}_r$),
effective tangential pressure ($p^{eff}_t$) have been shown graphically in Figs. 1 and 2. Here
$\rho^{eff}$, $p^{eff}_r$, $p^{eff}_t$ all are maximum at the centre with positive signature and decrease while
approach to the surface. We can measure the effective surface density and verify the radius
of the star from the cut on $r$-axis. These high values of the central
as well as the effective surface density clearly emphasize the fact that
our chosen stellar candidates are highly compact, and thus actually
represent themselves as strange quark star candidates~\cite{Ruderman1972,Glendenning1997,Herzog2011}.
On the other hand, plot of the anisotropic stress (Fig. 3) demonstates the physical stability of our model.

$\textbf{(ii) TOV equation:}$
In our model the plots (Fig. 4) for the generalized force condition (TOV
equation), show that our stellar model remains in static equilibrium
under the combined effect of four different forces, viz. hydrostatic
force ($F_h$), gravitational force ($F_g$), anisotropic force ($F_a$)
and the additional modified gravity force ($F_{mg}$). Here, the newly
added force, represented as the modified gravity force ($F_{mg}$) implies
the coupling between matter and geometry.

$\textbf{(iii) Energy conditions:}$
In our study, we have graphically represented (Fig. 5) that the variation
of different energy conditions, namely WEC, NEC, SEC and DEC satisfies
for the prescribed anisotropic fluid distribution consisting of strange
quark matter.

$\textbf{(iv) Stabilty of model:}$
Following Herrera's~\cite{Herrera1992} cracking condition, $v^2_{rs}$ and
$v^2_{ts}$ should lie between the limit 0 and 1. Figs.~\ref{vel.} and ~\ref{veld.} clearly show that
$v^2_{rs}$, $v^2_{ts}$ , $|v^2_{rs}-v^2_{ts}|$ remain in this limit, within the fluid distribution.
So from cracking concept and causality condition
~\cite{Herrera1992,Andreasson2009}, our model is physically reasonable
as well as potentially stable. In the left panel of Fig.~\ref{vel.}, $v^2_{rs}$ is not a constant, rather it
shows non-linearly decreasing nature with the increasing radii and
numerical value remains slightly smaller than $1/3$ for all the stars.
Here, putting $\chi=0$ in Eq.~(\ref{eq40}), we get back $v^2_{rs}=1/3$ which signifies
the constant value for radial sound speed in GR.

Variation of the EOS parameter for both the radial and tangential cases,
have been displayed in Fig.~\ref{eos.}. Within the fluid distribution, value of the EOS parameter
is always positive and less than 1, which is another evidence for the
stability of our proposed model. On the other hand, variations of adiabatic indices,
plotted in Fig.~\ref{adia.}, evidently show that
both $\Gamma_r$ and $\Gamma_t$ are greater than $4/3$ throughout the
stellar system, obeying Bondi's~\cite{Bondi1964} stable configuration
criteria.

In this model, static stability criteria privided by Harrison, Zel$'$dovich and Novikov
~\cite{Harrison1965,Zeldovich1971} is also satisfied (Eq.~(\ref{eq51})) for
different strange stars. Fig.~\ref{HZN.} also shows that $\frac{dM}{d\rho_c}>0$ always
within the stellar structure and reduces for higher radial value.

$\textbf{(v) Buchdahl Condition:}$ The effective mass function up to the surface
(i.e. the radius $r$) has been shown in Fig.~\ref{mass.}. This figure shows that
$M_{eff}(r)\rightarrow 0$ for $r\rightarrow0$ which emphasizes on
the regularity of $M_{eff}(r)$ at $r=0$, i.e., at the centre. In case of the spherically
symmetric, static and perfect fluid distribution, Buchdahl~\cite{Buchdahl1959}
established a condition for the mass and radius ratio, i.e., $\frac{2M}{\Re}\leq \frac{8}{9}$.
In our study, we consider three strange star candidates (Table 1), for which Buchdahl
condition~\cite{Buchdahl1959} is satisfied. Here $\frac{2M}{\Re}$ ratio exists
in the range $(0.29-0.35)$.

$\textbf{(vi) Compactness and surface redshift:}$ We have studied
three different strange star candidates, namely $PSR~J~1614-2230$, $Vela~X-1 $, $Cen~X-3$
whose mass and radius are provided in Table 1~\cite{Deb2016}. Variation
of the compactification factor w.r.t. $r/{\Re}$ has been presented in Fig.~\ref{cf.} where
the revealed features are highly reasonable for strange stars.
Here we find high surface redshift $(0.2-0.25)$ which establishes
that our model stars represent some possible candidates for strange stars which are
stable in their configuration.

In connection to stability, there are several research works available on modified gravity
on strange stars which provide stability though the problem of singularity
arises at the centre. On the other hand, few works do not satisfy all the stability
criteria, energy conditions, Buchdahl limit~\cite{Buchdahl1959} one at a time.
However, in the present study using the KB metric and the MIT bag model in modified gravity (i.e. $f(R,T)$),
our proposed model is completely free from any singularity and satisfies all the stability criteria.

As  a final concluding remark, the present study on
stellar model is nothing but the representative of highly dense
stars formed with strange quark matter and perfectly suitable for
investigating various features of strange stars. Besides that,
most fascinating fact is the effect of modified gravity on the bag constant $B_g$ as
shown in Table~\ref{Table3}. Due to the coupling between matter and geometry, there arises
a coupling term $\chi$ which effectively reduces the bag value as well as the
square of radial sound speed, which generally remains constant ($1/3$) in GR.
In every case, putting $\chi=0$, one can retrieve the results which perfectly
match the GR results.

\section*{Acknowledgement}
SR and FR are thankful to the Inter University Centre for
Astronomy and Astrophysics (IUCAA) for providing Visiting Associateship under
which a part of this work has been carried out. SR is also thankful to the Authority
of The Institute of Mathematical Sciences, Chennai, India for providing all
types of working facility and hospitality under Associateship scheme. SB is
thankful to DST-INSPIRE [~IF~160526] for financial support and all types of facilities
for continuing research work. We are grateful to the anonymous referee for several
useful suggestions which have enabled us to modify the manuscript substantially.

\end{document}